\documentclass[a4paper,longauth]{aa}

\usepackage{graphicx}
\usepackage{txfonts}
\usepackage{natbib}
\usepackage{longtable,lscape}
\usepackage{amssymb}
\usepackage{mathrsfs}
\usepackage{amsfonts}
\usepackage{url}
\usepackage{color}
\usepackage{hyphenat}
\usepackage{natbib}
\bibpunct{(}{)}{;}{a}{}{,} 

\usepackage{ textcomp }
\usepackage{ gensymb }
\usepackage{ragged2e}

\newcommand{\arcm}{'}
\newcommand{\parcm}{\overset{'}{.}}
\newcommand{\planck}{\textit{Planck }}
\newcommand{\Planck}{\textit{Planck }}

\usepackage{pdflscape}
\usepackage{amsfonts}
\usepackage{amssymb}
\usepackage{amsmath}
\usepackage{rotate}%
\usepackage{graphicx}
\usepackage{rotating}%
\usepackage{lscape}%
\usepackage{epstopdf}
\usepackage{epsfig}
\usepackage{longtable,booktabs}
\usepackage[normalem]{ulem}
\usepackage{textcomp} 
\usepackage[utf8]{inputenc}
\usepackage{adjustbox}
\usepackage{makecell}

\usepackage{threeparttablex}
\usepackage{xtab,booktabs}

\hypersetup{draft}

\begin{document}

\title{Optical validation and characterization of \planck PSZ2 sources at the Canary Islands observatories. II. Second year of LP15 observations}


\subtitle{}

\author{A.~Aguado-Barahona \inst{1,2}\and R.~Barrena \inst{1,2}\and
  A.~Streblyanska \inst{1,2} \and A.~Ferragamo \inst{1,2} \and
  J.A.~Rubiño-Martín \inst{1,2} \and {D.~Tramonte} \inst{3,1,2} \and
  {H.~Lietzen} \inst{4} }

\institute{Instituto de Astrofísica de Canarias, C/Vía Láctea s/n, E-38205 La
   				Laguna, Tenerife, Spain\\
              \email{aaguado@iac.es}
         \and
             Universidad de La Laguna, Departamento de Astrofísica, E-38206
				La Laguna, Tenerife, Spain
		 \and
		 	University of KwaZulu-Natal, Westville Campus, Private Bag X54001, Durban 4000, South Africa
		 \and
		 	Tartu Observatory, University of Tartu, Observatooriumi 1, 61602, Tõravere, Estonia
             }

   \date{}

\abstract
{The second legacy catalogue of Planck Sunyaev-Zeldovich (SZ) sources, hereafter
  PSZ2, provides the largest galaxy cluster sample selected by means of their SZ
  signature in a full sky survey. In order to fully characterise this PSZ2
  sample for cosmological studies, all the members should be validated and the
  physical properties of the clusters, including mass and redshift, should be
  derived. However, at the time of its publication roughly 21 per cent of the
  1653 PSZ2 members had no known counterpart at other wavelengths. }
{ Here, we present the second and last year of observations of our optical
  follow-up programme 128-MULTIPLE-16/15B (hereafter LP15), which has been
  developed with the aim of validating all the unidentified PSZ2 sources in the
  northern sky, with declination above $-15\degree$, and with no correspondence
  in the first Planck catalogue PSZ1. The description of the programme and the
  first year of observations were presented in Streblyanska et al. (2019).  }
{ The LP15 programme was awarded 44 observing nights, spread over two years in
  the Isaac Newton Telescope (INT), the Telescopio Nazionale Galileo (TNG) and
  the Gran Telescopio Canarias (GTC), all at Roque de los Muchachos Observatory
  (La Palma). Following the same methodology described in Streblyanska et
  al. (2019), at the end of the LP15 programme we performed deep optical imaging
  for more than 200 sources with the INT, and spectroscopy for almost 100
  sources with the TNG and GTC. We adopted a robust confirmation criteria based
  on velocity dispersion and richness estimations in order to carry out the
  final classification of the new galaxy clusters as the optical counterparts of
  the PSZ2 detections. }
{Here, we present the observations of the second year of LP15, as well as the
  final results of the programme. The full LP15 sample comprises 190 previously
  unidentified PSZ2 sources. Of those, 106 objects were studied in Streblyanska
  et al. (2019), while the remaining sample (except for 6 candidates) has been
  completed in the second year and it is discussed here. In addition to the LP15
  sample, in this paper we have studied 42 additional PSZ2 objects, which were
  originally validated as real clusters due to their matching with a WISE or
  PSZ1 counterpart, but they had no measured spectroscopic redshift. In total,
  we have confirmed the optical counterpart for 81 PSZ2 sources after the full
  LP15 programme, 55 of them with new spectroscopic information. Out of those 81 sources, 40 clusters are presented in this paper. After the LP15 observational
  programme the purity of the PSZ2 catalogue has increased from $76.7\,\%$
  originally to $86.2\,\%$. In addition, we study the possible reasons of having
  false detection, and we report a clear correlation between the number of
  unconfirmed sources and galactic thermal dust emission. }
{}

\titlerunning{Optical validation of \planck\ PSZ2. II.}
\maketitle
%

\section{Introduction}
	
Galaxy Clusters (GCs) are a very powerful tool to test the cosmological model
\citep{Allen11}. They have been used for cosmology studies since Zwicky's
discovery of dark matter in the Coma Cluster \citep{Zwicky33}, and thanks to the
recent all-sky surveys appearing during the last decade, their study has
increased significantly. In particular, the
\Planck\footnote{\Planck\ \url{http://www.esa.int/Planck} is a project of the
  European Space Agency (ESA) with instruments provided by two scientific
  consortia funded by ESA member states and led by Principal Investigators from
  France and Italy, telescope reflectors provided through a collaboration
  between ESA and a scientific consortium led and funded by Denmark, and
  additional contributions from NASA (USA).} satellite \citep{Planck14I}
provided for the first time the possibility for the detection of GCs on a
full-sky survey by means of the Sunyaev-Zeldovich (SZ) effect \citep{SZ72},
which is Cosmic Microwave Background (CMB) spectral distortion. As CMB photons
go through a GC, they can interact with the hot intra-cluster medium (ICM)
electrons via inverse Compton scattering. As a result of this interaction, the
CMB photons gain energy and the overall CMB spectrum is shifted towards higher
frequencies, producing a characteristic spectral dependence which can be used
for their detection.

In order to use the GCs surveys to constrain cosmological parameters
\citep{2018arXiv180706209P, 2016A&A...594A..24P}, it is important to obtain
unbiased measurements of the cluster mass and redshift. In particular, due to
the fact that the surface brightness of the SZ effect does not depend on
redshift, follow-up campaigns at other wavelengths are needed in order to
complement the SZ information. The detailed characterisation of
\planck SZ survey cluster sample is needed to improve cosmological
constraints from \planck survey alone, and also to obtain constraints from
future large galaxy cluster surveys, e.g. SRG/eROSITA cluster
survey \citep{eRosita}, whose spacecraft was successfully launched 
on July 13th and the first light of SRG/eROSITA telescope is planned to be obtained very soon.

The PSZ2 catalogue \citep{PSZ2} is the second \planck catalogue of
Sunyaev-Zeldovich sources derived from the full 29 months mission data. This
catalogue is based on the results from three cluster detection codes (MMF1, MMF3
and PwS) described in detail in \cite{PSZ1,PSZ2}. The PSZ2 was validated using
external X-ray, optical, SZ and near-IR data and contained, at the time of its
publication, 1653 detections of which 1203 were confirmed as actual galaxy
clusters, 1094 of them including redshifts. The PSZ2 catalogue contains all
objects found by at least one of the three detection algorithms with $S/N \geq
4.5$ for the SZ detection.
	
A first validation process was performed in \cite{PSZ2}. It began using a
cross-match with the PSZ1, continuing the search for possible counterparts in
the MCXC catalogue \citep{Piffaretti11} which is based on the \textit{ROSAT} All
Sky Survey \citep[RASS,][]{Voges99,Voges00} and  on the serendipitous \textit{ROSAT} and Einstein cluster catalogues, in the Sloan Digital Sky Survey
\citep[SDSS][]{York00}, in the redMaPPer catalogue \citep{Rykoff14}, in
NED\footnote{The NASA/IPAC Extragalactic Database (NED) is operated by the Jet
  Propulsion Laboratory, California Institute of Technology, under contract with
  the National Aeronautics and Space Administration} in similar follow-ups
\citep{PcolintXXVI, paper1} as the one presented in this work, in the AllWISE
mid-infrared source catalogue \citep{Cutri13} as well as in SZ catalogues such
as the catalogues obtain by the South Pole Telescope \citep[SPT,][]{Bleem15}, by
the Atacama Cosmology Telescope \citep[ACT,][]{Hasselfield13} and by direct
follow-up with the Arc-minute Micro-kelvin Interferometre
\citep[AMI,][]{Perrott15}.
   
This paper is the last of a series of papers where optical characterisation of
SZ sources was performed using the Canary Islands
Observatories. \citet{paper1,paper2} study PSZ1 sources thanks to the
observational programme {\tt ITP13-08}. Later, a second long-time programme was
granted {\tt 128-MULTIPLE-16/15B} (hereafter {\tt LP15} ) to study the PSZ2
sources. \citet[][hereafter Paper I]{paper4} and this paper are the result of
this last programme. The main motivation of these follow-up campaigns is to
identify and confirm optical cluster counterparts of unknown sources. We perform
photometric and spectroscopic observations in order to study the optical
richness and estimate velocity dispersion. We leave the mass estimates for
future works.

This paper is structured as follows. Sections \ref{sec:observations} and
\ref{sec:criteria} describe the observational programme carried out for this
work, including the instrumentation setups, the observational approach as well
as the criteria used to validate candidates as optical counterparts for SZ
sources. Section \ref{sec:2ndyear} presents the observations for the 2nd year of
this programme and details some special cases. In Section \ref{sec:wise} we
update information about already known counterpart validated using ALLWISE
\citep{Cutri13}. Section \ref{sec:fullpsz2} summarises the whole programme and
gives the final results for this work. Section \ref{sec:conclusions} presents
the conclusions.
	
We adopt $\Lambda$CDM cosmology with $\Omega_{\rm m} = 0.3075$,
$\Omega_{\Lambda} = 0.691$ and H$_0 = 67.74$\,km\,s$^{-1}$\,Mpc$^{-1}$.

\section{LP15 optical follow-up campaign}
\label{sec:observations}

\subsection{Sample definition and observational strategy}

The {\tt LP15} programme, the sample definition and the observing strategy are
described in detail in paper I. Here, we briefly summarise its basic
characteristics and present the second year of observations. As a reminder, the
main motivation of the programme is to carry out a systematic follow-up of the
complete set of PSZ2 cluster candidates in the northern sky, with no confirmed
counterparts at the moment of the catalogue publication. For LP15, we consider
only sources located at declination $> -15\degree$. As a reference, the full
PSZ2 catalogue contains a total set of 1003 sources in that region.

The {\tt LP15} sample is defined by all those sources in PSZ2 with declination
above $-15\degree$, which also have {\tt validation = -1} (i.e., no known
counterpart at the time of the publication of the catalogue), and {\tt PSZ1 =
  -1} (i.e. no matching detection in the PSZ1). This corresponds to 190 targets
in total, 106 of which were already discussed in paper I.

During the second year of the programme, covering the semesters 2016B and 2017A,
we continued with the observations of the remaining {\tt LP15} sample. As for the
previous year, the second year of observations were carried out at the Roque de
los Muchachos Observatory (ORM) located at La Palma island (Spain) between
August 2016 and August 2017. The three telescopes used in this work are: a) the
2.5\,m Isaac Newton Telescope (INT) operated by the Isaac Newton Group of
Telescopes; b) the 3.6\,m Italian Telescopio Nazionale Galileo (TNG) operated by
the Galileo Galilei Foundation of the INAF (Istituto Nazionale di Astrofisica);
and c) the 10.4\,m Gran Telescopio Canarias (GTC) operated by the Instituto de
Astrofísica de Canarias (IAC). More information about their instruments and
technical features were presented in Table~1 of Paper I, as well as
information about the observing nights and the number of candidates observed.

Our follow-up programme is structured as follows. The first step is to select a
target for deep-imaging observations and if galaxy over-densities are found,
confirm the cluster using spectroscopy. The photometric observations are carried
out in the INT using three broadband Sloan filters ($g^\prime$, $r^\prime$,
$i^\prime$). Based on colour combinations of these filters, we are able to
estimate photometric redshifts \citep{paper1} of possible members of the cluster
up to $z\sim 0.8$. In the second step, candidates are definitely confirmed using
multi-object spectroscopy (MOS) either in the TNG or in GTC, depending on the
photometric estimation of the redshift of the candidate. Due to the greater
collecting area of the GTC, we use it for distant clusters ($z>0.35$), while the
TNG for the closest ones ($z<0.35$). The final step is to validate the
candidates using the confirmation criteria explained in
Sect.~\ref{sec:criteria}.

\subsection{Imaging and spectroscopic observation and data reduction}
\label{sec:imagespecdata}

The complete technical description of the telescopes and instrument set-ups used
during {\tt LP15} programme and information on corresponding data reduction is
detailed in Paper I. Here, we briefly summarise the information of the
imaging and spectroscopic observations and data reduction performed during the
second year of the programme.
		
Optical observations were carried out during multiple runs between August 2016
and August 2017. Deep images were obtained using the Wide-Field Camera (WFC)
installed in the 2.5\,m in INT. Spectroscopic observations were obtained using
the multi-object spectrographs DOLORES@TNG and OSIRIS@GTC. The spectroscopic
data from DOLORES was retrieved during multiple runs between November 2016 and
August 2017 and the data from OSIRIS was acquired in service mode between
September 2016 and August 2017.

All images and spectra were reduced using {\tt IRAF}\footnote{{\tt IRAF}
  (\url{http://iraf.noao.edu/}) is distributed by the National Optical Astronomy
  Observatories, which are operated by the Association of Universities for
  Research in Astronomy, Inc., under the cooperative agreement with the National
  Science Foundation.} standard routines. The astrometry was performed using the
{\tt images.imcoords} task and the USNO B1.0 catalogue \citep{Monet03} as
reference obtaining a final accuracy of {\it rms} $\sim 0.2\arcsec$ across the
full field of view. The photometric calibration is referred to SDSS
photometry. In the case of fields with SDSS coverage, we performed the direct
cross-correlation with the SDSS photometric data. Using {\tt SExtractor}
\citep{Bertin96} programme in single-image mode, we created individual source
catalogues for each band. All catalogues were then merged in the final one with
a search radius of $1\arcsec$.

Standard spectroscopic data reduction included sky subtraction, extraction of
spectra, cosmic rays rejection, and wavelength calibration (using specific
arcs). Our final reduced spectra have typical $S/N \sim 5$ per pixel (at around
$6000\,\AA$) for galaxies with magnitudes $r^\prime \sim 20.5$ and $21.7$
observed with the TNG and GTC, respectively.

\begin{figure*}[h!]
\centering 
\includegraphics[width=\columnwidth]{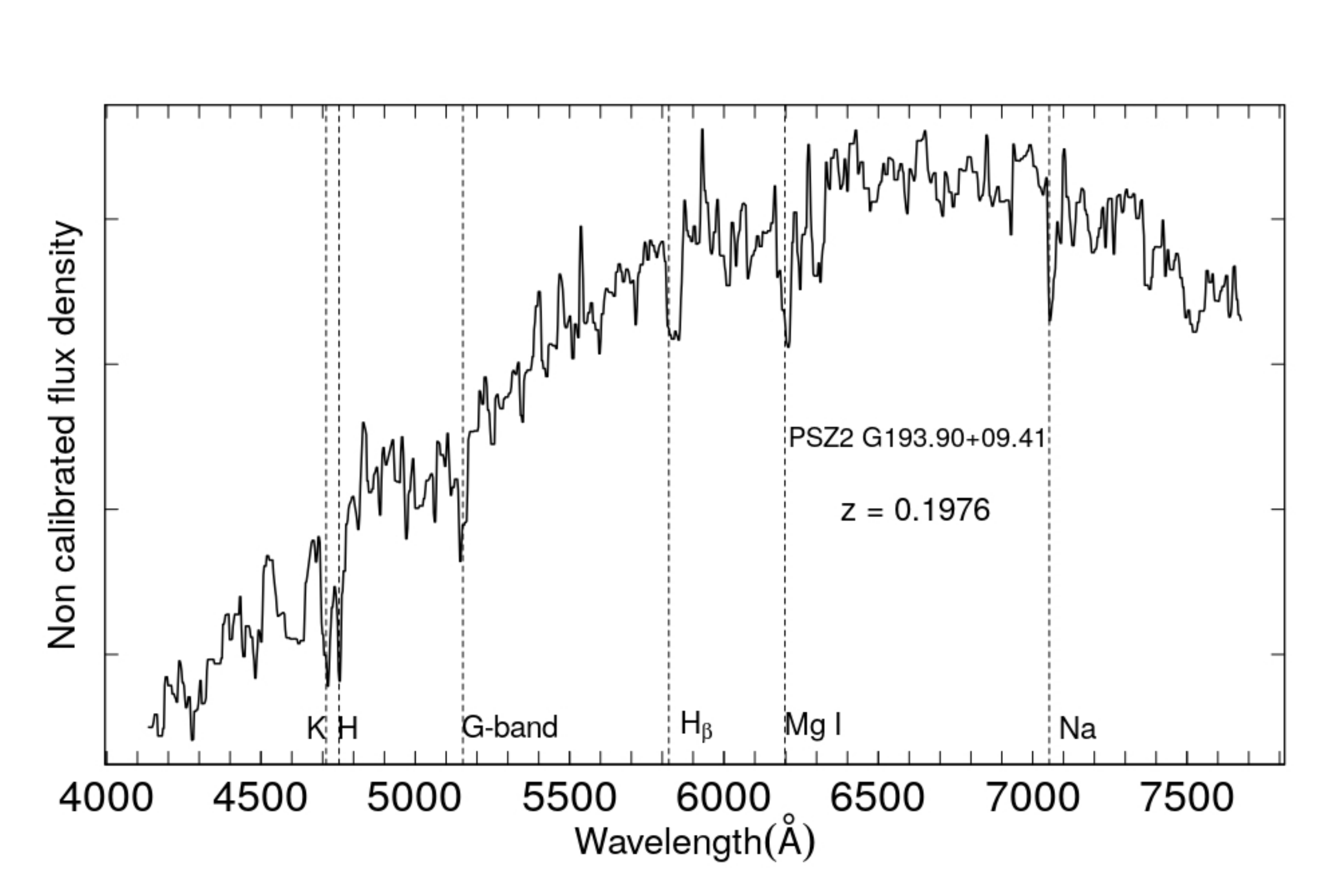}
\includegraphics[width=\columnwidth]{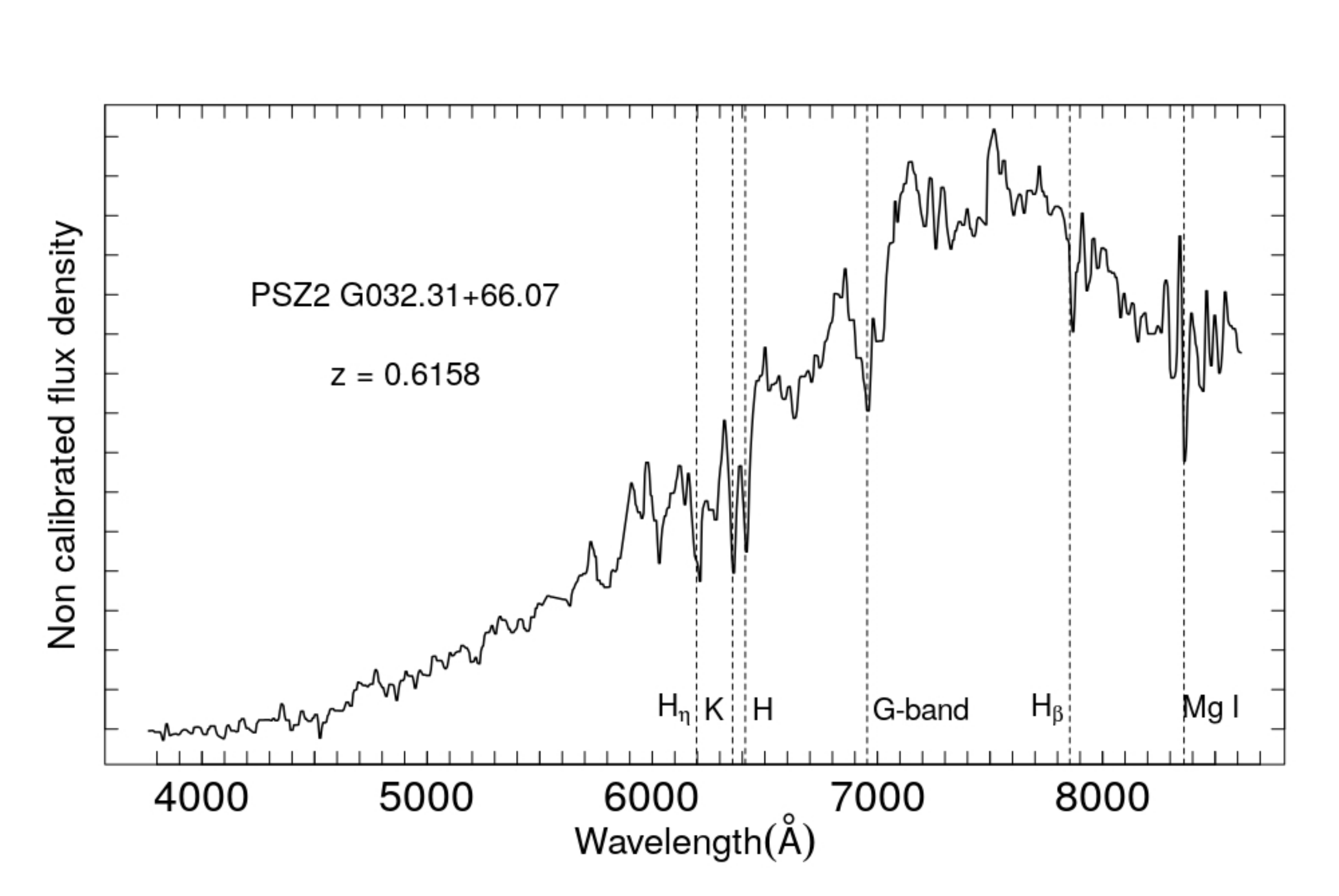}
\caption{Example of the spectra obtained with TNG/DOLORES (left panel) and
  GTC/OSIRIS (right panel) for two luminous galaxy members in the PSZ2
  G193.90$+$09.41 and PSZ2 G032.31$+$66.07 clusters, at $z=0.1976$ and
  $z=0.6158$, respectively. Dashed lines correspond to the wavelength of the
  absorption features identified in each spectrum at the redshift of the
  clusters. Flux density is plotted in arbitrary units.}
\label{fig1}
\end{figure*}

We estimated radial velocities using task {\tt RVSAO}\footnote{RVSAO was
  developed at the Smithsonian Astrophysical Observatory Telescope Data
  Center.}.  Figure \ref{fig1} shows two examples of spectra for low and high-z
galaxies obtained at the TNG and GTC. Thanks to the capabilities of the MOS
observations, we were able to retrieve typically 20 members per cluster with
which to estimate the mean redshift and velocity dispersion of the systems. The
cluster redshift is taken to be the mean value of the galaxy members
retrieved. The galaxies are considered members only if they showed radial
velocities of $\pm 2500$\,km\,s$^{-1}$ in rest frame with respect to the mean
velocity of the system. Then, we follow an iterative method considering galaxies
as members if their radial velocity is less than 2.5 times the velocity
dispersion away from the cluster mean velocity. We follow this procedure in
order to minimise the contamination of interlopers. The velocity dispersion and
mass of the clusters will be published in a future work.
	
The broad band images used to carry out this work has been already included in
the Virtual Observatory (VO) collection for public access. In the near future,
the photometric and spectroscopic catalogues will be also available through this
platform.

\section{Cluster identification and validation criteria}
\label{sec:criteria}

Here we describe the methodology used to validate a cluster candidate as the
optical counterpart of a PSZ2 target. This procedure is similar to the one
adopted in paper I, and it is an extension of the methodology applied in
\citet{paper1,paper2}. Compared to other methods in the literature, we have
improved the validation criteria by including visual inspection and comparison
between the RGB images, and also the Compton $y$-maps
    \citep{Planck15XXII}, making photometric redshift estimates and analysing the
red sequence \citep[hereafter RS][]{Gladders00} using colour-magnitude
diagrams. We have also performed a richness study considering galaxy counts in
clusters. For approximately 30\,$\%$ of our sample we have performed
spectroscopic confirmation by estimating the velocity dispersion of the
candidates.

\subsection{Photometric analysis}
\label{sec:cri_photo}

The first step in our validation process is to visually inspect the deep RGB
images and compare them with the Compton $y$-map of the SZ source. Having as
reference the nominal \planck pointing coordinates, we look for galaxy
over-densities around $5\arcm$ radius, which is 2.5 times the \planck mean
position error \citep{PSZ2}. However, for a few cases, when a possible
counterpart is found further than $5\arcm$, we perform a more detailed
analysis. It must be clear that in this first step we identify galaxy
over-densities in the field of view but not everyone of them is associated with
the SZ source. The next steps are performed in order to study this possible
association.
	
We use colour-magnitude diagrams to identify likely clusters members and fit the
RS to make an estimate of the photometric redshift using the methodology
explained in \cite{paper2}, Sect.~3 and in \cite{paper1}, Sect.~4.2. To do so,
we use $(g^\prime - r^\prime)$ and $(r^\prime -i^\prime)$ colours and consider
galaxies within $\pm 0.05$ from the Brightest Cluster Galaxy (BCG) colour.

\begin{figure} 
  \centering
  \includegraphics[width=\columnwidth]{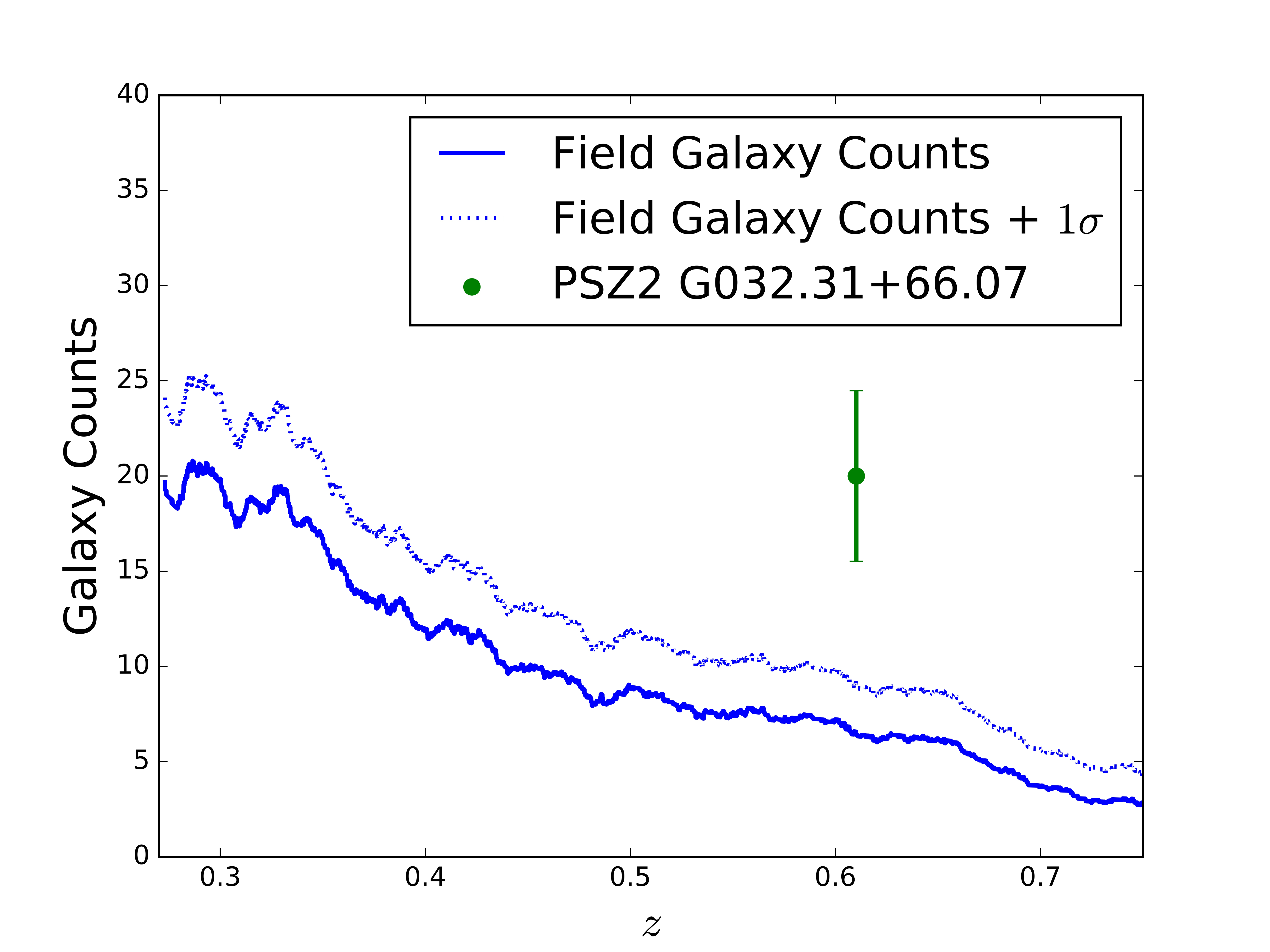}
\caption{Galaxy counts as a function of redshift in the field of the
  spectroscopically confirmed cluster PSZ2 G032.31$+$66.07. The galaxy counts
  for this particular cluster and its 1-$\sigma$ error bars are shown in
  green. The blue line represents the galaxy counts outside 1\,Mpc region from
  the optical centre of the cluster and the dashed blue line represents
  1$\sigma$ uncertainty above the latter.}
\label{figrich}
\end{figure}

The SZ clusters of the \planck catalogue are expected to be massive structures
($\ga 5 \times 10^{14}$\,M$_\odot$) \citep[see][]{PSZ1} and therefore, they
should present rich galaxy cluster populations. In order to discard the low mass
systems from the real massive counterparts, we define a richness parameter
($R_{0}$) and its significance above the background level $\sigma_{R}$. The
detailed description of our richness calculations is given in Paper I. Briefly, for a given cluster, we count likely cluster members
(assumed as galaxies within the RS $\pm 0.15$, in colour) showing $r'$-magnitudes
in the range $[m_{r^{\prime}}^{\star}-1, m_{r^{\prime}}^{\star}+1.5]$ within the
1\,Mpc from the cluster centre. For this task, we consider as a BCG the most
luminous galaxy of the identified likely cluster members. In order to obtain the
background subtracted richness and decontaminate from the galaxy field
contribution, we compute galaxy counts outside the 1\,Mpc region and using the
same restrictions in colour and magnitude as explained before, thereby creating
a field galaxy sample or local background for each cluster ($R_{f}$). This field
sample is scaled to the 1\,Mpc area and subtracted from the cluster counts to
obtain a statistical estimate of the number of galaxies for each cluster, the
so-called richness ($R_{cor} \equiv R_{0}-R_{f}$). We based our confirmation on
the significance above the background level $\sigma_{R}$ which is computed as
$R_{cor} / \sqrt{R_{f}}$. For a better understanding see Fig. \ref{figrich}
which shows galaxy counts as a function of redshift in the field of the
confirmed cluster PSZ2 G032.31$+$66.07. The initial value found for this cluster
was $R_{0} = 20.0$, and for the field at the redshift of the cluster was $R_{f}
= 6.4$, yielding to $R_{cor} = 13.6$ and $\sigma_{R} = 5.4$.

\subsection{Spectroscopic analysis}
\label{sec:cri_spec}

We performed spectroscopic observations for approximately 30\,$\%$ of our
sample, and include public data from the SDSS spectroscopic archive. We built
radial velocity catalogues for each cluster and obtained the velocity dispersion
in order to distinguish between poor and massive systems.
	
\planck systems must have masses $M_{500} \gtrsim 10^{14}$\,M$_\odot$ at low
redshift ($z < 0.2$) and $M_{500} \gtrsim 2 \times 10^{14}$\,M$_\odot$ at higher
redshift \cite[see][Fig.4]{vanderBurg16}. Using the scaling relations M$_{500} -
\sigma_{v}$ from \cite{Munari13} and $M_{500}$--$M_{200}$ from \cite{Komatsu11},
those values of mass lead to a velocity dispersion of $\sigma_{v} >
500$\,km\,s$^{-1}$ for $z < 0.2$ and $\sigma_{v} > 650$\,km\,s$^{-1}$ for $z
\geq 0.2$. We adopt here the same criteria used in previous validation studies
of \planck clusters \citep{paper1,paper2,paper4}. The velocity dispersion of the
candidates will be presented in a future work.

\subsection{Confirmation criteria adopted in this work}
\label{sec:cri_cri}

\begin{table}[t]
\centering
\caption{Vadation criteria adopted to confirm or reject candidates associated to the SZ emission.}
\begin{tabular}{c c c c}
\noalign{\vskip 3pt\hrule\vskip 5pt}
Flag & Spectroscopy & $\sigma_{v}$ (km s$^{-1}$) & $\sigma_{R}$ \cr
\noalign{\vskip 3pt\hrule\vskip 5pt}
1  &  YES  & $> 500$ km s$^{-1}$  (z $<$ 0.2)  & $> 1.5$     \cr 
   &       & $> 650$ km s$^{-1}$  (z $>$ 0.2)  & $> 1.5$     \cr
\noalign{\vskip 3pt\hrule\vskip 5pt}
2  &  NO   &        $-$                        & $> 1.5$ \cr
\noalign{\vskip 3pt\hrule\vskip 5pt}
3  &  YES  & $< 500$ km s$^{-1}$  (z $<$ 0.2)  & $> 1.5$     \cr 
   &       & $< 650$ km s$^{-1}$  (z $>$ 0.2)  & $> 1.5$     \cr
   &  NO   &        $-$                        & $< 1.5$ \cr
\noalign{\vskip 3pt\hrule\vskip 5pt}
ND &  $-$  &        $-$                        & $-$     \cr
\hline
\end{tabular}


\label{tab:cri}
\end{table}
	
Table~\ref{tab:cri} summarises the set of criteria adopted in this work in order to
confirm or reject a cluster candidate as the optical counterpart of the SZ
signal. These are the very same criteria used in Paper I, and
provide a classification of the candidates according to four possible values of
a {\tt Flag}. Values of {\tt Flag}$=1$ or 2 correspond to validated clusters;
{\tt Flag}$=3$ corresponds to clusters located along the line of sight of the
\planck signal but possibly not associated with the SZ emission; and ND refer to
non-detection.
For those cases where we have enough spectroscopic information to provide an
estimate of $\sigma_{v}$, if that value is found to be above the corresponding
threshold, we validate the candidate with {\tt Flag} 1. However, if $\sigma_{v}$
is below the threshold, we assume that the system has a low mass and it is
probably not linked to the SZ emission, being the candidates classified as {\tt
  Flag} 3. In the case that no spectroscopic information is available, or if we
cannot estimate of the velocity dispersion due to an insufficient number of
galaxy members (less than $5$ members), we validate the candidates using the
richness estimate. Systems showing a $\sigma_{R} > 1.5$ are validated
photometrically, but waiting for a definitive spectroscopic confirmation. These
systems are classified with {\tt Flag} 2. Clusters with {\tt Flag} 3 represent
very poor systems ($\sigma_{R} < 1.5$) with no spectroscopic information. The ND
(non-detection) flag is used for those SZ candidates where we found no galaxy
over-density in the optical images. We also consider the criterion that a
\planck cluster must be placed within 5$\arcm$ radius from the nominal pointing
because it represents 2.5 times the mean position error with respect to the SZ
peak emission. Nevertheless, this criterion can be modulated due to the wide
range of uncertainties in the position error in the PSZ2 catalogue, and the
different shapes of the $y$-maps. The cases that do not match the validation
criteria but are positively confirmed will be discussed in
Sect.~\ref{sec:notes}.

\section{LP15 sample: 2nd year of observations}
\label{sec:2ndyear}

\begin{table*} 
\begin{center}
\caption[]{\label{tab:LP15summary}Summary information of the long-term {\tt
    LP15} programme. For each year of the programme, we show the total number of
  observed candidates (column 2), the total number of validated clusters (column
  3), and the fraction of those with spectroscopic measurements (column 4). For
  completeness, columns 5--8 also include the classification of the candidates
  according to our validation criteria described in Table~\ref{tab:cri}. Note that
  validated clusters (column 3) are those with flags 1 or 2. The full {\tt LP15} sample contained 190 candidates. Thus, there are still 6 additional objected to be studied (see text for details).}
  
\begin{tabular}[h]{cccccccc}
\hline
Year   &  Observed &   val & spec & {\tt Flag 1} & {\tt Flag 2} & {\tt Flag 3} & ND \\
\hline
   1   &       106 &    41 &   34 &    31 & 10  &   8   & 57 \\
   2   &        78 &    40 &   22 &    18 & 22  &   6   & 32 \\
\hline
TOTAL: &       184 &    81 &   56 &    49 & 32  &   14  & 89 \\ 
\hline
\end{tabular}
\end{center}
\small 
\end{table*}

Table~\ref{tab:LP15summary} summarises the basic information of programme LP15
after the two years of observations, concerning the characterisation of the {\tt LP15}
sample. The results for year one of the programme were already discussed in Paper I.
Here, table~\ref{tab:LP15} presents the results for 78 PSZ2 galaxy cluster
candidates studied in this optical follow-up during the second and last
year. The table is organised as follows. Columns 1, 2 and 3 are the official ID
number, the \planck Name and the SZ signal-to-noise ratio, respectively, as they
appear in the PSZ2 catalogue. Columns 4 and 5 are the J2000 coordinates of the
BCG when present, otherwise geometrical centre of the cluster is
provided. Column 6 is the distance between \planck and the optical centre
reported in this work. Columns 7 and 8 present the spectroscopic information
when available: the mean spectroscopic redshift of the cluster and/or the BCG,
and the number of spectroscopic members retrieved. Columns 9, 10 and 11 provide
the photometric information: the photometric redshift, the estimation of the
richness and the value $\sigma_{R}$ as explained in
Sect.~\ref{sec:cri_photo}. Column 12 lists the cluster classification following
the {\tt Flag} system described in Sect.~\ref{sec:cri_cri}. Finally, column 13
is reserved for special comments or notes about the individual candidates.

Following the confirmation criteria explained in Sect.~\ref{sec:cri_cri}, we
find that 37 of our candidates have a single optical counterpart, and one
additional is classified as double detection. We classify a source as double
detection when we find two or more over-densities around the SZ emission peak
that might contribute to this emission. The ones validated with a single optical
counterpart are classified as follows: 17 as {\tt Flag} 1 and 22 as {\tt Flag} 2. In addition, we find 32 non-detection, flagged as ND, and six systems not associated with the corresponding SZ source ({\tt Flag} 3). This means a total of 38 PSZ2 sources remaining unconfirmed. 
	
We have partially focused our work using SDSS DR12 data to confirm PSZ2 clusters
classified by \cite{Streblyanska18} as `potentially associated' with the SZ
emission. We have obtained the redshift and the velocity dispersion for six of
the photometrically confirmed clusters and we have re-confirmed six clusters
using our own deep INT imaging data classifying them as {\tt Flag} 1 and 2,
respectively. From the `potentially associated' sub-sample of
\cite{Streblyanska18} we have confirmed five as {\tt Flag} 1. Here, we
invalidate the PSZ2 G328.96$+$71.97, confirmed by \cite{Streblyanska18}. New
SDSS DR14 data reveals that the counterpart proposed by the authors is part of a
larger system whose BCG is 34$\parcm$6 away from the \planck SZ pointing. This
system will be discussed in detail in Section \ref{sec:notes}.

\begin{figure}
\centering
\includegraphics[width=\columnwidth]{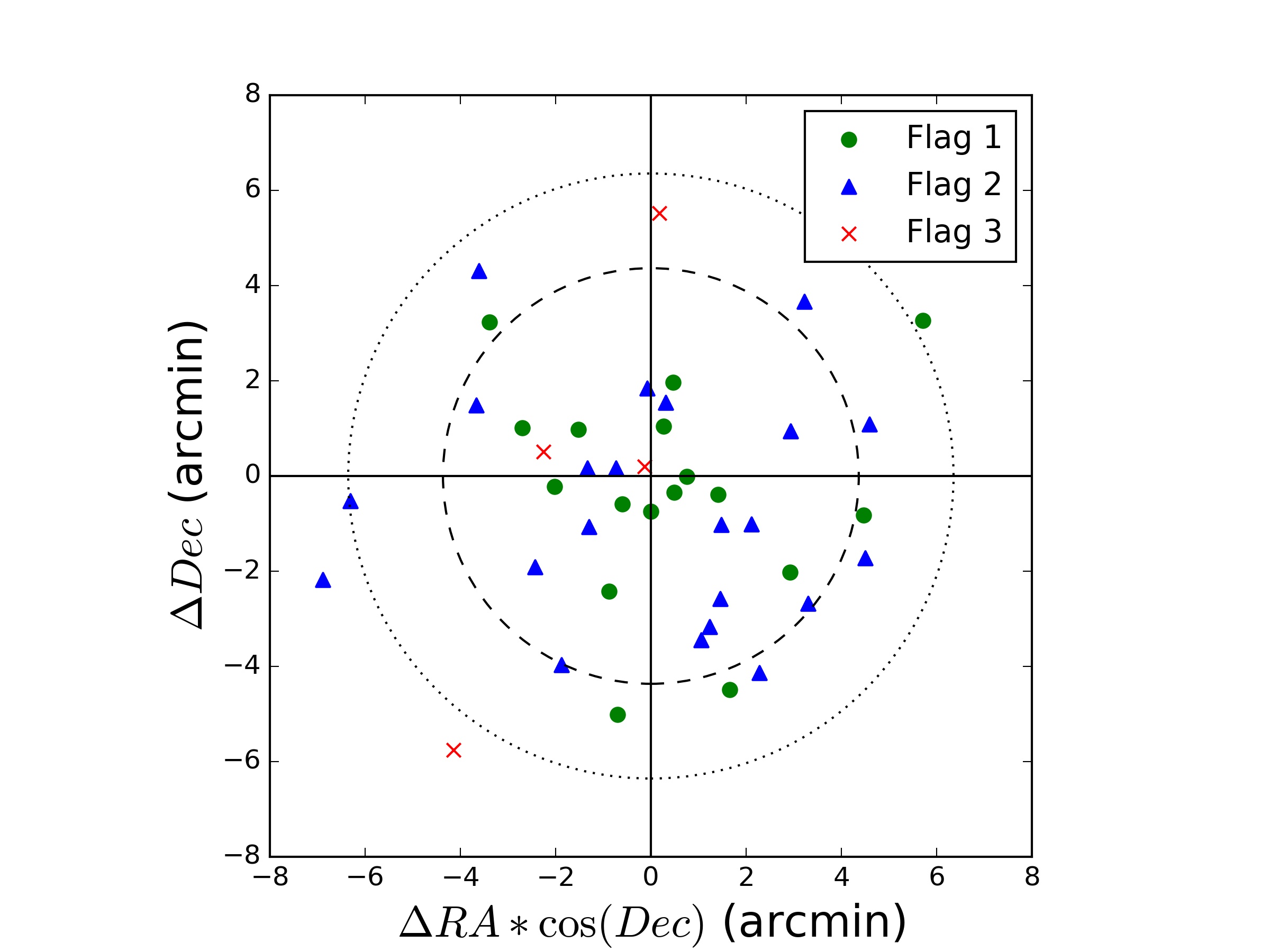}
\caption{Spatial distribution of the optical counterparts centres with respect
  to the nominal centre in the \planck PSZ2 catalogue. Green dots, blue
  triangles and red crosses correspond to clusters classified with {\tt Flag} 1,
  {\tt Flag} 2 and {\tt Flag} 3, respectively. Dashed and dotted lines show the
  regions enclosing 68\,$\%$ and 95\,$\%$ of the confirmed clusters, flagged as
  `1' and `2', respectively}
\label{Figoff}
\end{figure}

Figure~\ref{Figoff} represents the spatial distribution of the optical
counterpart centre with respect to the nominal centre in the \planck PSZ2
catalogue for the clusters in our sample flagged as `1',`2' and `3'. The median
of the offsets of the clusters validated is $3\parcm 10$ which is in agreement
with the median of the position error computed from the PSZ2 catalogue ($2\parcm
43$). The mean values are also in agreement ($3.3$ and $2.6$) but they are more
sensitive to cases which their position error is too high. We find 25 (35)
clusters within $4\parcm 46$ ($6\parcm 38$) representing the 68\,$\%$ (95\,$\%$)
of the confirmed clusters.

\begin{figure}
\centering
\includegraphics[width=\columnwidth]{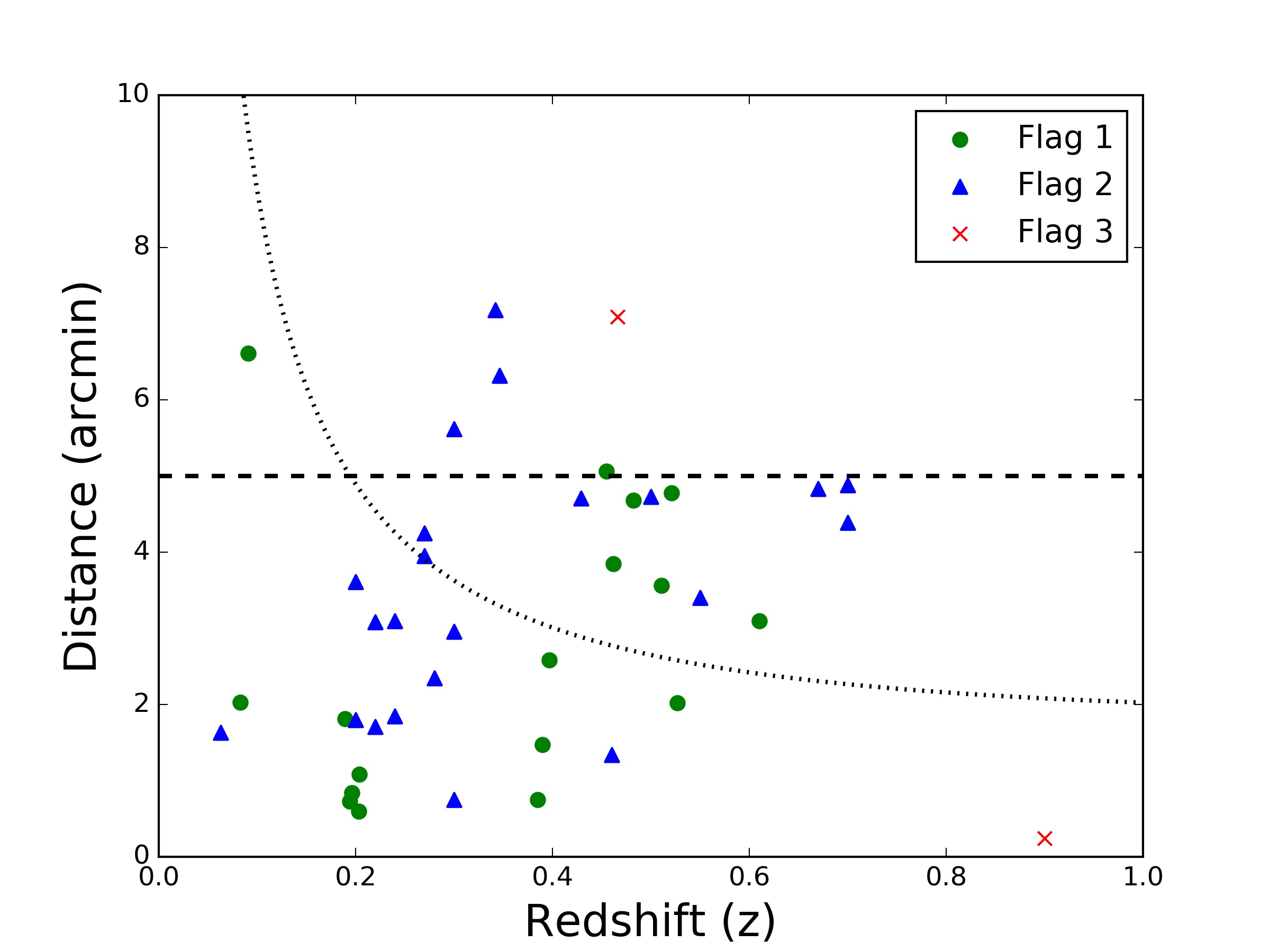}
\caption{Cluster optical centre offsets relative to their \planck SZ position as
  a function of cluster redshift. The dashed horizontal line at $5\arcm$ shows
  the maximum offset expected for a \planck SZ detection. The dotted line
  corresponds to the angle subtended by 1\,Mpc in projection at the
  corresponding redshift. Symbols used are the same as in Fig.~\ref{Figoff}.}
\label{Figdz}
\end{figure}
		
Figure~\ref{Figdz} shows the cluster optical centre offsets relative to their
\planck SZ position as a function of cluster redshift. Although the limit in
distance is $5\arcm$ (see Sect.~\ref{sec:criteria}), there are some confirmed
sources that exceed that value due to different reasons such as the high
position error in the PSZ2 catalogue or elongated $y$-map contour around the SZ
emission peak (see Sect. \ref{sec:notes}).

\begin{figure}
\centering
\includegraphics[width=\columnwidth]{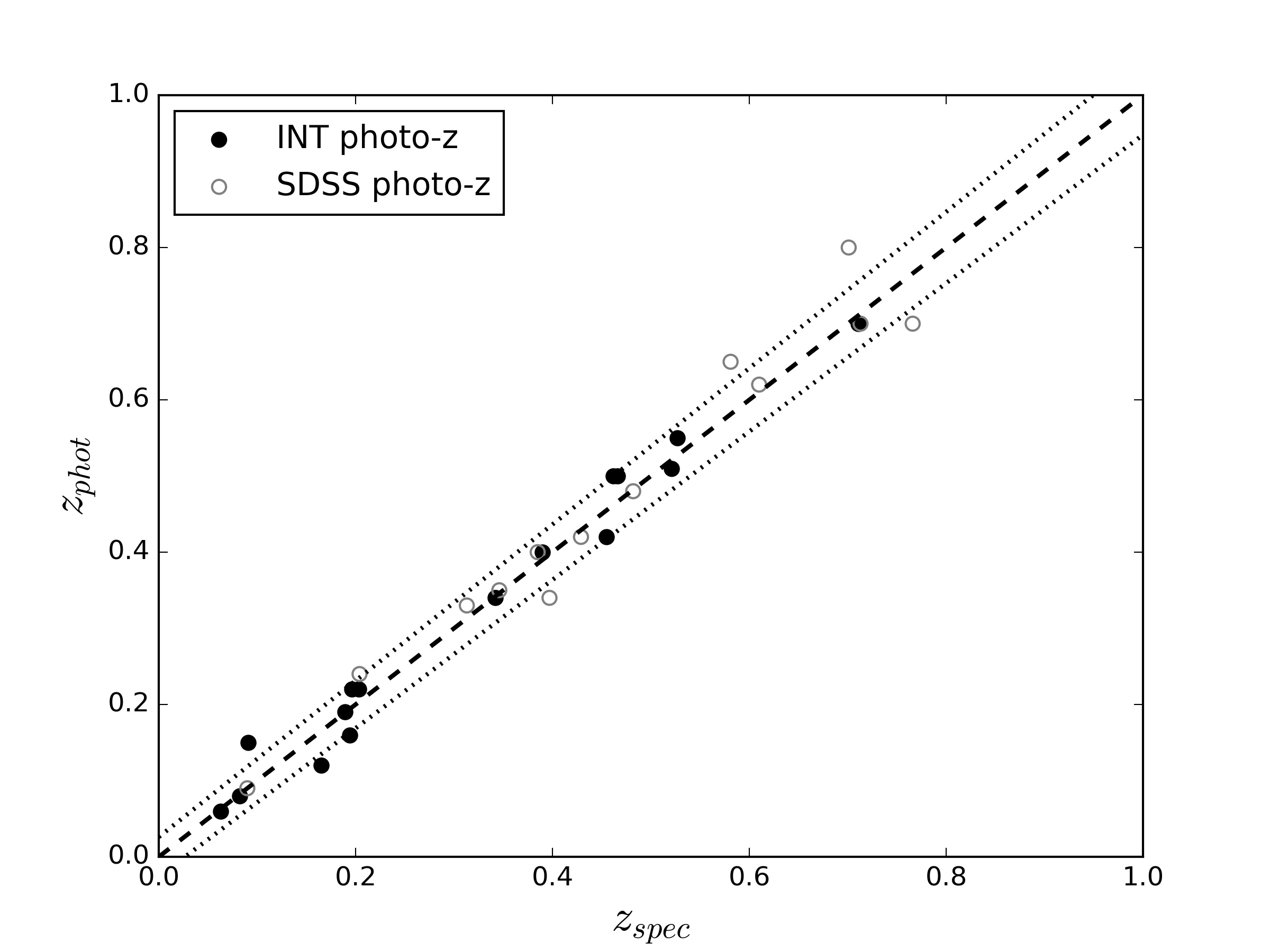}
\caption{Comparison between photometric and spectroscopic redshift estimations.
  Dashed line represents the 1:1 relation while dotted lines show the
  photometric redshift error $\delta z/(1+z) = 0.026$.}
\label{Figzz}
\end{figure}	

Figure~\ref{Figzz} presents a comparison between photometric and spectroscopic
redshift estimations. Photometric redshifts were obtained as explained in
Sect.~\ref{sec:observations}. We show every observation carried out during the
second year of the {\tt LP15}, including the ones explained in
Sect.~\ref{sec:wise}. This study yields a mean photometric redshift error of
$\delta z/(1+z) = 0.026$.

Finally, we note that there are six objects in the {\tt LP15} sample that have
not been observed during the programme. One of them, PSZ2 G186.50$-$13.45, was
already validated in \cite{Streblyanska18}, with a photometric redshift of
$z_{\rm phot}=0.25$. According to our validation criterion, this case would
correspond to a {\tt Flag}$=2$.

A dedicated proposal has been submitted and approved to complete the
observations for the remaining five objects: PSZ2 G023.05+20.52, PSZ2
G092.34+14.22, PSZ2 G206.55-43.22, PSZ2 G210.37-37.00 and PSZ2
G247.14+25.88. These observations will be conducted in July 2019.


\begin{landscape} 

\setlength\LTleft{10pt}
\tiny

\begin{table}
\caption[]{List of 78 PSZ2 cluster candidates analysed in this paper.} \label{tab:LP15}
\tiny
\begin{tabular}{@{\extracolsep{\fill}}p{1cm} c c c c c c c c c c c p{4.5cm} @{}}
\hline \hline
\noalign{\smallskip}
\,   & \multicolumn{6}{c}{Position (J2000)} \\
\cline{4-5}
\noalign{\smallskip}
ID$^{1}$&  Planck Name & SZ SNR & R. A. & Decl.& Dist.($\arcm$) & $<z_{\rm spec}>$ ; $z_{\rm spec,BCG}$&$N_{\rm spec}$&$z_{\rm phot}$&$R_{\rm cor}$&$\sigma_{\rm R}$&{\tt Flag} & Notes$^{2}$ \\

\noalign{\smallskip}
\hline 
\noalign{\smallskip}


 115$^{a,b}$      & PSZ2 G032.31$+$66.07 & 5.14 & 14 37 23.35 & $+$24 24 21.70 & 3.10 & 0.610 ;\hspace{0.4cm}$-$\hspace{0.25cm} & 38  & 0.62$\pm$0.05 & 13.6$\pm$3.7 &  5.4 &  1 &  (1), sub-structured \\
 194              & PSZ2 G048.47$+$34.86 & 5.74 &     $-$     &        $-$     &  $-$ &      $-$       & $-$ &      $-$      &      $-$     &  $-$ & ND &  \\ 
 242              & PSZ2 G058.31$+$41.96 & 4.54 &     $-$     &        $-$     &  $-$ &      $-$       & $-$ &      $-$      &      $-$     &  $-$ & ND &  \\
 421-A$^{c}$      & PSZ2 G092.69$+$59.92 & 4.90 & 14 26 03.78 & $+$51 14 18.50 & 3.85 & 0.462 ; 0.4568 & 25  & 0.50$\pm$0.05 & 11.6$\pm$3.4 &  4.3 &  1 & (1,2)\\
 421-B$^{c}$      &                      &      & 14 26 13.10 & $+$51 11 53.17 & 4.42 & 0.844 ;\hspace{0.4cm}$-$\hspace{0.25cm} &  5  &      $-$      &      $-$     &  $-$ &  3 & (3)\\
 424$^{b}$        & PSZ2 G093.41$-$16.26 & 4.59 & 22 24 07.25 & $+$37 58 30.46 & 3.10 &      $-$       & $-$ & 0.24$\pm$0.03 & 40.6$\pm$6.4 &  7.9 &  2 &  WHL J222407.2$+$375831\\ 
 432$^{a,b}$      & PSZ2 G094.31$-$11.31 & 4.72 & 22 12 56.10 & $+$42 35 46.34 & 1.08 & 0.204 ;\hspace{0.4cm}$-$\hspace{0.25cm} & 27  & 0.24$\pm$0.03 &      $-$     &  $-$ &  1 & \\
 500              & PSZ2 G104.52$+$39.39 & 4.60 & 15 58 38.88 & $+$70 27 24.20 & 5.62 &      $-$       & $-$ & 0.30$\pm$0.04 & 16.8$\pm$4.1 &  7.8 &  2 & \\
 511              & PSZ2 G105.94$-$16.14 & 4.62 &             &                &  $-$ &      $-$       & $-$ &      $-$      &      $-$     &  $-$ & ND & \\
 545              & PSZ2 G112.54$+$59.53 & 5.37 &     $-$     &        $-$     &  $-$ &      $-$       & $-$ &      $-$      &      $-$     &  $-$ & ND &\\
 546$^{c}$        & PSZ2 G112.69$+$33.37 & 4.63 & 16 19 49.39 & $+$79 06 24.49 & 4.78 & 0.521 ; 0.5194 & 15  & 0.51$\pm$0.03 &      $-$     &  $-$ &  1 &  (1),WHL J161949.3$+$790624 \\
 592              & PSZ2 G120.75$+$25.39 & 4.69 &     $-$     &        $-$     &  $-$ &      $-$       & $-$ &      $-$      &      $-$     &  $-$ & ND & \\
 600              & PSZ2 G122.81$+$24.74 & 4.60 &     $-$     &        $-$     &  $-$ &      $-$       & $-$ &      $-$      &      $-$     &  $-$ & ND & \\
 613              & PSZ2 G125.25$+$33.33 & 5.38 & 11 41 11.26 & $+$83 27 38.91 & 1.80 &      $-$       & $-$ & 0.20$\pm$0.03 & 21.0$\pm$4.6 &  4.1 &  2 & \\
 616              & PSZ2 G125.41$+$27.95 & 4.76 &     $-$     &        $-$     &  $-$ &      $-$       & $-$ &      $-$      &      $-$     &  $-$ & ND & \\
 620$^{b}$        & PSZ2 G125.84$-$18.72 & 5.30 & 01 06 55.65 & $+$44 04 25.72 & 1.81 & 0.189 ;\hspace{0.4cm}$-$\hspace{0.25cm} & 46  & 0.19$\pm$0.01 &      $-$     &  $-$ &  1 &  WHL J010709.2$+$440918 \\
 624$^{b}$        & PSZ2 G126.36$-$19.11 & 5.01 & 01 09 19.57 & $+$43 37 40.41 & 0.60 & 0.203 ; 0.2007  & 22  & 0.22$\pm$0.01 &      $-$     &  $-$ &  1 &  WHL J010919.5$+$433741 \\
 627              & PSZ2 G126.62$-$53.42 & 4.55 &     $-$     &        $-$     &  $-$ &      $-$       & $-$ &      $-$      &      $-$     &  $-$ & ND & \\
 628$^{b}$        & PSZ2 G126.72$-$21.03 & 4.68 & 01 10 27.91 & $+$41 40 57.27 & 0.84 & 0.196 ;\hspace{0.4cm}$-$\hspace{0.25cm} &  9  & 0.22$\pm$0.03 &      $-$     &  $-$ &  1 &  WHL J011025.0$+$414119 \\
 640              & PSZ2 G129.99$-$22.42 & 4.55 &     $-$     &        $-$     &  $-$ &      $-$       & $-$ &      $-$      &      $-$     &  $-$ & ND & \\
 644-A$^{b}$      & PSZ2 G130.64$+$37.16 & 4.80 & 10 47 45.54 & $+$77 59 56.67 & 2.82 & 0.473 ; 0.4722 & 14  & 0.44$\pm$0.03 & 16.7$\pm$4.1 &  5.0 &  1 &  WHL J104745.5$+$775957 \\
 644-B$^{b}$      &                      &      & 10 46 29.31 & $+$78 07 44.06 & 6.07 &                & $-$ & 0.24$\pm$0.02 & 49.3$\pm$7.0 & 11.9 &  2 & (1), WHL J104629.2$+$780744 \\
 646$^{b}$        & PSZ2 G131.15$-$14.72 & 5.37 & 01 38 42.22 & $+$47 22 35.27 & 1.71 &      $-$       & $-$ & 0.22$\pm$0.03 & 50.2$\pm$7.1 & 11.3 &  2 &  WHL J013846.1$+$472236 \\
 647              & PSZ2 G131.19$+$14.48 & 4.80 &     $-$     &        $-$     &  $-$ &      $-$       & $-$ &      $-$      &      $-$     &  $-$ & ND & \\
 648              & PSZ2 G131.27$-$25.82 & 4.50 &     $-$     &        $-$     &  $-$ &      $-$       & $-$ &      $-$      &      $-$     &  $-$ & ND & \\
 667$^{c}$        & PSZ2 G136.02$-$47.15 & 4.64 & 01 28 23.61 & $+$14 41 13.60 & 7.09 & 0.466 ; 0.4648 &  8  & 0.50$\pm$0.03 & 13.1$\pm$3.6 &  4.2 &  3 &  WHL J012823.6$+$144114\\
 700              & PSZ2 G143.90$+$25.06 & 4.91 &     $-$     &        $-$     &  $-$ &      $-$       & $-$ &      $-$      &      $-$     &  $-$ & ND & \\
 712$^{a}$        & PSZ2 G146.10$-$55.55 & 4.67 & 01 42 46.53 & $+$04 59 48.72 & 4.71 & 0.429 ;\hspace{0.4cm}$-$\hspace{0.25cm} &  1  & 0.42$\pm$0.04 & 10.8$\pm$3.3 &  3.0 &  2 & \\
 713              & PSZ2 G146.13$+$40.97 & 4.90 & 09 40 17.10 & $+$66 24 02.56 & 7.18 & 0.342 ; 0.3379 &  4  & 0.34$\pm$0.04 &  7.0$\pm$2.6 &  3.5 &  2 & \\
 717              & PSZ2 G146.88$+$17.13 & 6.13 &     $-$     &        $-$     &  $-$ &      $-$       & $-$ &      $-$      &      $-$     &  $-$ & ND & \\
 720              & PSZ2 G147.17$+$42.67 & 4.92 & 09 50 01.152 & $+$64 55 29.52 & 1.34 &\hspace{0.25cm}$-$\hspace{0.3cm}; 0.4401 &   1 & 0.46$\pm$0.04 &  7.4$\pm$2.7 &  3.1 &  2 & (4)\\
 727              & PSZ2 G149.73$+$24.49 & 4.52 &     $-$     &        $-$     &  $-$ &      $-$       & $-$ &      $-$      &      $-$     &  $-$ & ND & \\
 732$^{a,b}$      & PSZ2 G150.64$-$14.21 & 4.68 & 03 17 04.20 & $+$40 41 33.22 & 3.08 &      $-$       & $-$ & 0.22$\pm$0.02 & 26.1$\pm$5.1 &  8.2 &  2 &  WHL J031704.2$+$404133 \\
 739$^{a,c}$      & PSZ2 G152.40$+$75.00 & 4.70 & 12 13 19.17 & $+$39 46 26.84 & 5.06 & 0.455 ;\hspace{0.4cm}$-$\hspace{0.25cm} &  9  & 0.42$\pm$0.03 & 31.3$\pm$5.6 &  9.0 &  1 &  (1),WHL J121319.2$+$394627\\
 740              & PSZ2 G152.47$+$42.11 & 4.81 & 09 29 52.64 & $+$61 39 40.00 & 0.24 & 0.900 ;\hspace{0.4cm}$-$\hspace{0.25cm} &  6  &      $-$      &      $-$     &  $-$ &  3 & \\
 746              & PSZ2 G153.68$+$36.96 & 5.07 &     $-$     &        $-$     &  $-$ &      $-$       & $-$ &      $-$      &      $-$     &  $-$ & ND & \\
 747              & PSZ2 G153.80$+$33.79 & 4.52 &     $-$     &        $-$     &  $-$ &      $-$       & $-$ &      $-$      &      $-$     &  $-$ & ND & \\
 754              & PSZ2 G156.24$+$22.32 & 4.79 & 06 45 02.14 & $+$59 27 13.30 & 0.75 &      $-$       & $-$ & 0.30$\pm$0.05 &      $-$     &  $-$ &  2 & \\
 769              & PSZ2 G160.94$+$44.85 & 4.98 &     $-$     &        $-$     &  $-$ &      $-$       & $-$ &      $-$      &      $-$     &  $-$ & ND & \\
 780              & PSZ2 G163.89$+$11.55 & 4.78 &     $-$     &        $-$     &  $-$ &      $-$       & $-$ &      $-$      &      $-$     &  $-$ & ND & \\
 788              & PSZ2 G165.39$+$09.22 & 5.60 & 05 48 09.37 & $+$46 04 41.41 & 4.39 &      $-$       & $-$ & 0.70$\pm$0.10 & 15.1$\pm$3.9 &  5.4 &  2 &  Clearly sub-structured, second centre at 05:48:10.56 $+$46:07:18.38 \\
 789              & PSZ2 G165.41$+$25.93 & 4.51 & 07 23 27.93 & $+$52 07 32.70 & 4.83 &      $-$       & $-$ & 0.67$\pm$0.04 &  4.3$\pm$2.1 &  1.8 &  2 &  \\
 797              & PSZ2 G166.56$-$17.69 & 4.76 & 04 04 53.39 & $+$28 18 31.85 & 3.44 &      $-$       & $-$ & 0.70$\pm$0.07 &  3.3$\pm$1.8 &  1.4 & ND &  \\
 799              & PSZ2 G167.43$-$53.67 & 4.65 &     $-$     &        $-$     &  $-$ &      $-$       & $-$ &      $-$      &      $-$     &  $-$ & ND & \\
 812$^{a,c}$      & PSZ2 G171.48$+$16.17 & 4.75 & 06 38 00.94 & $+$43 50 57.20 & 0.75 & 0.385 ; 0.3881 & 25  & 0.40$\pm$0.05 & 15.9$\pm$4.0 &  2.7 &  1 &  WHL J063743.6$+$434859\\
 820              & PSZ2 G173.76$+$22.92 & 5.80 & 07 17 26.66 & $+$44 05 00.28 & 1.63 & 0.063 ; 0.0652 &  2  & 0.06$\pm$0.02 &      $-$     &  $-$ &  2 & (5) \\
 831$^{a}$        & PSZ2 G177.03$+$32.64 & 4.93 & 08 13 08.56 & $+$43 13 53.07 & 3.56 & 0.511 ;\hspace{0.4cm}$-$\hspace{0.25cm} &  9  &      $-$      &      $-$     &  $-$ &  1 & (1)\\
 835              & PSZ2 G179.33$-$22.22 & 5.02 &     $-$     &        $-$     &  $-$ &      $-$       & $-$ &      $-$      &      $-$     &  $-$ & ND & \\
 836$^{a,b}$      & PSZ2 G179.45$-$43.92 & 4.54 & 03 19 18.34 & $+$02 05 35.60 & 2.58 & 0.397 ; 0.4005 & 23  & 0.34$\pm$0.03 & 19.1$\pm$4.4 &  5.8 &  1 &  WHL J031918.3$+$020535 \\
 849              & PSZ2 G183.32$-$31.51 & 4.56 & 04 05 20.11 & $+$07 51 26.07 & 2.31 &      $-$       & $-$ & 0.55$\pm$0.10 &  1.8$\pm$1.3 &  0.9 &  3 & \\
\hline

\end{tabular}
\end{table}
\end{landscape}

\addtocounter{table}{-1}
\begin{landscape}
\setlength\LTleft{10pt}

\begin{table}
\caption[]{Continue.} \label{tab:LP15}
\tiny
\begin{tabular}{@{\extracolsep{\fill}}p{1cm} c c c c c c c c c c c p{4.5cm} @{}}
\hline \hline
\noalign{\smallskip}
\,   & \multicolumn{6}{c}{Position (J2000)} \\
\cline{4-5}
\noalign{\smallskip}
ID$^{1}$&  Planck Name & SZ SNR & R. A. & Decl.& Dist.($\arcm$) & $<z_{\rm spec}>$ ; $z_{\rm spec,BCG}$&$N_{\rm spec}$&$z_{\rm phot}$&$R_{\rm cor}$&$\sigma_{\rm R}$&{\tt Flag} & \multicolumn{1}{c}{Notes$^{2}$} \\

\noalign{\smallskip}
\hline 
\noalign{\smallskip}

 852              & PSZ2 G183.92$+$16.36 & 4.97 & 07 01 30.22 & $+$32 54 51.20 & 6.61 & 0.091 ; 0.0914 & 18  & 0.15$\pm$0.03 & 30.3$\pm$5.5 &  4.4 &  1 &  ABELL 567 \\
 859              & PSZ2 G185.68$+$09.82 & 5.18 & 06 37 14.93 & $+$28 38 02.80 & 1.47 & 0.390 ; 0.3897 & 39  & 0.40$\pm$0.05 &      $-$     &  $-$ &  1 & \\
 860              & PSZ2 G185.72$-$32.23 & 5.12 &     $-$     &        $-$     &  $-$ &      $-$       & $-$ &      $-$      &      $-$     &  $-$ & ND & \\
 878              & PSZ2 G191.57$+$58.88 & 5.17 &     $-$     &        $-$     &  $-$ &      $-$       & $-$ &      $-$      &      $-$     &  $-$ & ND & \\
 887              & PSZ2 G193.90$+$09.41 & 5.06 & 06 51 11.80 & $+$21 08 10.16 & 0.73 & 0.194 ; 0.1936 & 29  & 0.16$\pm$0.03 & 23.5$\pm$4.8 &  3.0 &  1 & \\
 912              & PSZ2 G201.20$-$42.83 & 4.70 &     $-$     &        $-$     &  $-$ &      $-$       & $-$ &      $-$      &      $-$     &  $-$ & ND & \\
 916$^{a,c,d}$    & PSZ2 G202.61$-$26.26 & 4.87 & 04 59 50.17 & $-$03 16 47.52 & 5.52 &      $-$       & $-$ & 0.23$\pm$0.03 &  8.5$\pm$2.9 &  4.6 &  3 & WHL J045950.2-031647\\
 917$^{a,c}$      & PSZ2 G202.66$+$66.98 & 4.63 & 11 07 30.90 & $+$28 51 01.20 & 4.68 & 0.482 ; 0.4814 & 20  & 0.48$\pm$0.04 & 11.9$\pm$3.4 &  5.3 &  1 & WHL J110730+285101\\
 920$^{a,c,d}$    & PSZ2 G203.32$+$08.91 & 5.15 & 07 05 56.53 & $+$12 30 33.66 & 4.25 &      $-$       & $-$ & 0.27$\pm$0.03 &  9.1$\pm$3.0 &  4.6 &  2 & WHL J070556.5$+$123034\\
 921$^{a}$        & PSZ2 G203.71$+$50.82 & 4.65 & 09 55 15.56 & $+$26 19 37.70 & 2.03 & 0.082 ;\hspace{0.4cm}$-$\hspace{0.25cm} & 22 &      $-$      &      $-$     &  $-$ &  1 & (1)\\
 952              & PSZ2 G210.71$+$63.08 & 7.37 &     $-$     &        $-$     &  $-$ &      $-$       & $-$ &      $-$      &      $-$     &  $-$ & ND & \\
 953              & PSZ2 G210.78$-$36.25 & 6.32 &     $-$     &        $-$     &  $-$ &      $-$       & $-$ &      $-$      &      $-$     &  $-$ & ND & \\
 982              & PSZ2 G218.58$+$08.71 & 4.63 & 07 32 40.27 & $-$01 03 21.55 & 3.40 &      $-$       & $-$ & 0.55$\pm$0.05 & 30.7$\pm$5.5 &  7.6 &  2 & 1RXS J073246.4$-$010205\\
1018              & PSZ2 G226.15$+$09.02 & 4.66 & 07 47 58.81 & $-$07 29 22.70 & 2.96 &      $-$       & $-$ & 0.30$\pm$0.05 & 28.9$\pm$5.4 & 10.2 &  2 & \\
1023              & PSZ2 G227.30$+$09.00 & 4.62 & 07 50 15.74 & $-$08 24 32.56 & 1.85 &      $-$       & $-$ & 0.24$\pm$0.03 &      $-$     &  $-$ &  2 & 1RXS J075020.3$-$082605\\
1049$^{a,b}$      & PSZ2 G231.41$+$77.48 & 4.54 & 12 00 26.54 & $+$22 34 19.55 & 6.32 & 0.346 ; 0.3469 &  2  & 0.35$\pm$0.03 &      $-$     &  $-$ &  2 & WHL J120026.5$+$223420 \\
1054              & PSZ2 G232.27$+$12.59 & 4.52 & 08 12 39.17 & $-$10 52 02.90 & 2.35 &      $-$       & $-$ & 0.28$\pm$0.04 & 35.3$\pm$5.9 & 10.3 &  2 & \\
1062              & PSZ2 G233.46$+$25.46 & 4.79 &     $-$     &        $-$     &  $-$ &      $-$       & $-$ &      $-$      &      $-$     &  $-$ & ND & \\
1074$^{a,b}$      & PSZ2 G237.68$+$57.83 & 5.36 & 10 53 17.80 & $+$10 52 37.13 & 4.88 &      $-$       & $-$ & 0.70$\pm$0.05 &      $-$     &  $-$ &  2 & (3)\\
1095              & PSZ2 G241.98$+$19.56 & 4.51 & 08 58 04.54 & $-$14 43 01.87 & 3.95 &      $-$       & $-$ & 0.27$\pm$0.04 & 33.1$\pm$5.7 &  8.3 &  2 & \\
1151              & PSZ2 G252.45$+$73.44 & 5.57 &     $-$     &        $-$     &  $-$ &      $-$       & $-$ &      $-$      &      $-$     &  $-$ & ND & \\
1162              & PSZ2 G253.95$+$39.12 & 4.66 &     $-$     &        $-$     &  $-$ &      $-$       & $-$ &      $-$      &      $-$     &  $-$ & ND & \\
1168              & PSZ2 G254.52$+$62.52 & 4.85 &     $-$     &        $-$     &  $-$ &      $-$       & $-$ &      $-$      &      $-$     &  $-$ & ND & \\
1219              & PSZ2 G263.96$+$40.64 & 4.58 &     $-$     &        $-$     &  $-$ &      $-$       & $-$ &      $-$      &      $-$     &  $-$ & ND & \\
1262$^{a,b}$      & PSZ2 G271.53$+$36.41 & 5.19 & 11 05 19.71 & $-$19 59 15.61 & 4.73 &      $-$       & $-$ & 0.50$\pm$0.03 &      $-$     &  $-$ &  2 &  WHL J110519.6$-$195852\\
1493              & PSZ2 G316.43$+$54.02 & 5.18 & 13 23 14.77 & $-$07 58 49.20 & 2.02 & 0.527 ; 0.5325 & 27  & 0.55$\pm$0.05 & 12.7$\pm$3.6 &  3.3 &  1 &  Sub-structured\\
1510$^{a,c}$      & PSZ2 G320.94$+$83.69 & 7.32 & 13 00 05.74 & $+$21 01 28.29 & 7.00 & 0.461 ; 0.4612 &  5  & 0.45$\pm$0.04 &      $-$     &  $-$ &  3 & \\
1513              & PSZ2 G321.94$+$75.57 & 4.66 &     $-$     &        $-$     &      &      $-$       & $-$ &      $-$      &      $-$     &  $-$ & ND & \\
1532              & PSZ2 G325.19$+$49.12 & 4.62 & 13 49 55.18 & $-$11 15 24.44 & 3.61 &      $-$       & $-$ & 0.20$\pm$0.03 & 28.1$\pm$5.3 &  6.5 &  2 &  WHY J135000.5$-$111724\\
1548$^{a,b}$      & PSZ2 G328.96$+$71.97 & 5.85 & 13 23 02.10 & $+$11 01 32.12 &18.03 & 0.090 ; 0.0937 & 94  & 0.09$\pm$0.01 &      $-$     &  $-$ &  3 & \\

\noalign{\smallskip}
\hline

\end{tabular}

\end{table}

\begin{tablenotes}[flushleft]
\tiny
\item[1] $^1$ SZ targets identified with the ID followed by an A or B label indicate the presence of multiple counterparts.
\item[2] $^2$ References. (1) \cite{Burenin17}, (2) \cite{Rykoff14}, (3) \cite{Burenin18}, (4) \cite{Zaznobin19}, (5) \cite{Boada18} 
\item[a] $^a$ Photometric and/or spectroscopic redshift obtained from SDSS DR14 data.
\item[b] $^b$ Already confirmed in \cite{Streblyanska18}
\item[c] $^c$ Classified as "potentially associated" in \cite{Streblyanska18}
\item[d] $^d$ Richness study from PAN-STARRS
\end{tablenotes}


\end{landscape}

\subsection{Notes on individual objects}
\label{sec:notes}

PSZ2 G058.31$+$41.96 This candidate is flagged as a non-detection due to a
bright star located near the \planck pointing. The star prevents photometric
measurements of this region, and thus, we are unable to visually identify an
over-density of galaxies. Despite this problem, we cannot identify visually
any over-density of galaxies in the region.

PSZ2 G104.52$+$39.39 The distance from the optical centre and the \planck
nominal pointing is $5\parcm 62$. Nevertheless, we validate this cluster with
{\tt Flag} 2 because the MILCA $y$-map contours are elongated along the line
that links both optical and \planck centres. Besides this fact, the position
error in the \planck catalogue is too high ($5\parcm 40$) compared to the
nominal one ($ 2\parcm 43 $).

\begin{figure}
\centering
\includegraphics[width=\columnwidth]{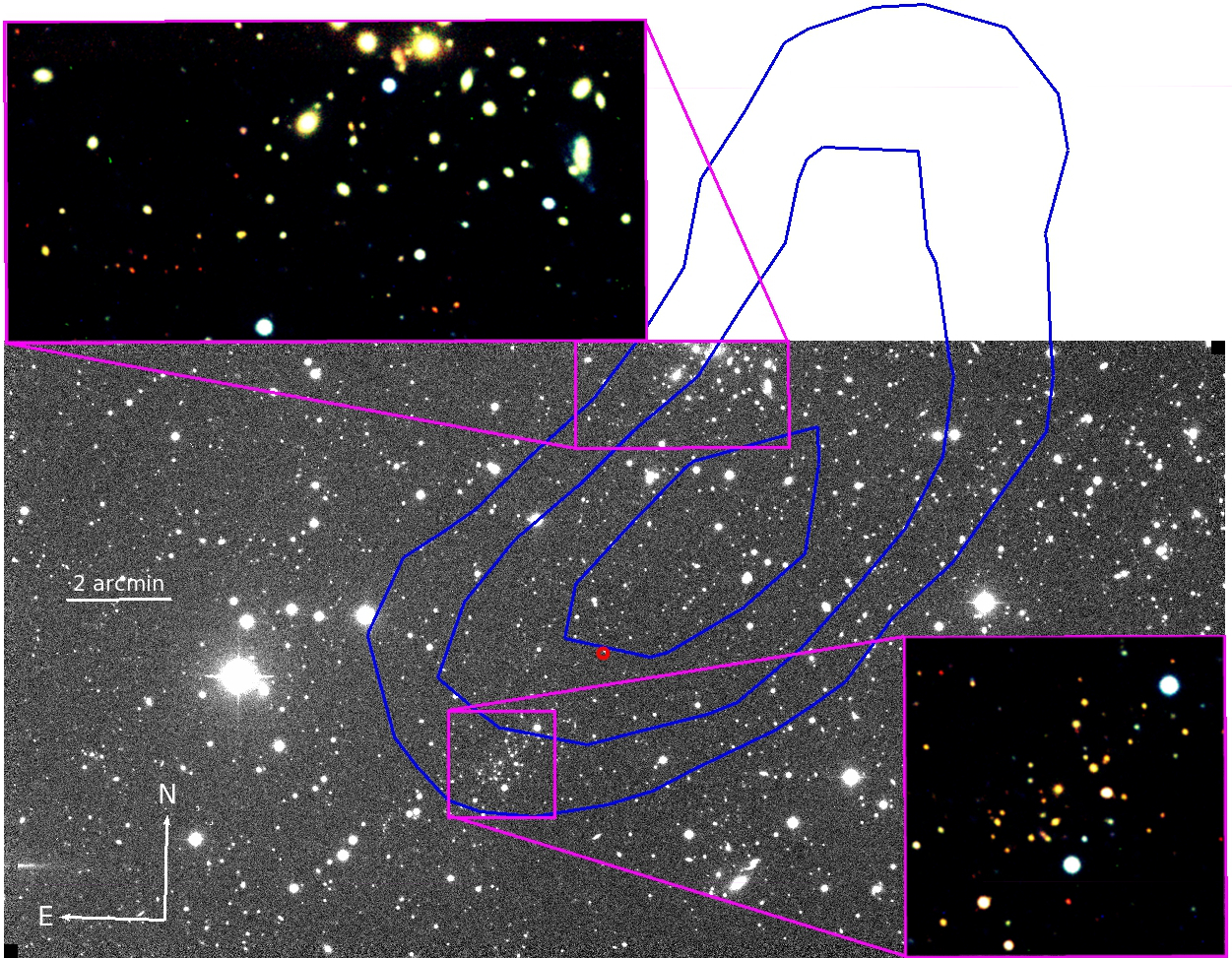}
\caption{Compton $y$-map superimposed on the INT $r'$-band of the PSZ2
  G130.64$+$37.16. Blue contours correspond to the $4$, $4.6$ and
  $5.2\times10^{-6}$ levels of the Compton $y$-map in this area. The nominal SZ
  pointing (red) is clearly closer to the cluster named 644-A (zoomed in the
  lower-right region) which presents 14 spectroscopic members and shows a
  velocity dispersion close to $1000$\,km\,s$^{-1}$; the cluster named 644-B
  (zoomed in the upper-left region) is $6\parcm 07$ away from the SZ centre but
  the MILCA $y$-map shows that the contours are elongated along this counterpart
  location which is twice as rich as the 644-A.}
\label{Fig643}
\end{figure}

PSZ2 G130.64$+$37.16 This candidate has two optical counterparts validated
already in \cite{Streblyanska18} and one of them in \cite{Burenin17}.
This is a tricky case as it is shown in
Fig. \ref{Fig643}. The nominal SZ pointing is clearly closer to the cluster
named 644-A which presents 14 spectroscopic members and shows a velocity
dispersion close to $1000$\,km\,s$^{-1}$; the cluster named 644-B is $6\parcm
07$ away from the SZ centre but the MILCA $y$-map shows that the contours are
elongated along this counterpart location which is twice as rich as the
644-A. This is also a case where the position error in the \planck catalogue is
$5\parcm 38$ which is more than twice the mean position error.

PSZ2 G146.13$+$40.97 The optical centre of the proposed counterpart (its BCG) is
at $7\parcm 18$ away from the Planck SZ pointing, which is affected by a
position error of $5\parcm 89$. The y-map contours present a very irregular
shape, maybe due to galactic dust contamination around this region. We estimate
a richness of $\sigma_{R}=3.5$ for this system and we find four cluster members
at $z_{\rm spec} = 0.342$ in the SDSS DR14 spectroscopic sample. So, we classify
this counterpart with {\tt Flag}$=2$. The ultimate confirmation will be obtained
using MOS observations.
	
PSZ2 G152.47$+$42.11 We find a possible cluster counterpart at $z_{\rm spec} =
0.900$. The deepness of our images makes it impossible to estimate the richness
at this redshift. However, we have observed this system spectroscopically and we
have found six cluster members. From these six galaxies, we obtain a very
low-velocity dispersion ($<400$\,km\,s$^{-1}$), revealing a low-mass galaxy
system. Thus, in this case, we classify this optical counterpart with {\tt
  Flag}$=3$.
	
PSZ2 G156.24$+$22.32 This region encloses two very bright stars, making it very
difficult to obtain accurate photometry or richness. However, by eye inspection,
we identify a cluster showing a galaxy population with coherent colours. A
detailed study of the photometry of some individual likely members and the BCG
reveals a $z_{\rm phot}=0.30$. In addition, the $y$-map contours present a very
regular profile centred on this system. For all these reasons, we classify this
system with {\tt Flag}$=2$.

PSZ2 G177.03$+$32.64 \cite{Burenin17} reports a counterpart for this
candidate at $z\sim0.28$. We have analysed this over-density finding seven galaxies
with spectroscopic redshifts in the SDSS archive. Four of these galaxies are
more than 4~Mpc away from the \planck centre and the velocity dispersion
accounting for the seven galaxies is less than $300$\,km\,s$^{-1}$. For this reason,
we present here only one counterpart at $z_{\rm spec} = 0.511$ whose velocity
dispersion, calculated using 9 members is approximately $1000$\,km\,s$^{-1}$.
	
PSZ2 G183.92+16.36 The distance between the BCG of this cluster and the \planck
pointing is $6\parcm 61$, which is only 0.67\,Mpc at the redshift of the
cluster, $z_{\rm spec} = 0.091$ (see Fig \ref{Figdz}). We perform multi-object
spectroscopy and retrieve 18 cluster members, showing a $\sigma_{v} \sim
650$\,km\,s$^{-1}$. In addition, this cluster is known as Abell 567, which is
well consolidated by other observations in the past \citep{Abell89}. Therefore,
we confirm Abell 567 as the counterpart of this SZ source, classifying it as
{\tt Flag}$=1$.

PSZ2 G202.61$-$26.26 and PSZ2 G203.32$+$08.91 Both candidates were analysed
using PANSTARRS photometric data \citep{PANSTARRS}. Both systems are rich but
the optical centre of the first one is more than $5 \arcm$ away from the \planck
nominal pointing, so classified as {\tt Flag 3}.
	
PSZ2 G227.30$+$09.00 This is an SZ source placed at very low galactic latitude, so
a large number of stars are crowding this field. For this reason, we were not
able to compute the richness: the galaxies of the background are partially
masked by the foreground stars. However, this system presents X-ray emission and
has been catalogued as 1RXS J075020.3-082605 in the ROSAT survey. So, we
classify this source as {\tt Flag}$=2$.
	
PSZ2 G237.68$+$57.83 This cluster has been already validated by
\cite{Streblyanska18} using SDSS data. Here, we confirm this association using
the INT images. Despite we are not able to perform a richness estimation at this
redshift, some individual likely cluster members show photometry in agreement
with a $z_{\rm phot} = 0.70\pm0.05$. We also find two additional over-densities
at (RA=10:53:35.55, Dec=+10:43:45.71) and (RA=10:53:59.602,
Dec=+10:46:38.23). However they are at over $> 10\arcm$ distance from the SZ
coordinates and thus, probably not contributing to the SZ signal. Therefore we
validate PSZ2 G237.68+57.83 as a single counterpart at $z_{\rm phot} =0 .70$.

PSZ2 G271.53$+$36.41 This candidate was confirmed photometrically in
\cite{Streblyanska18} as a double detection. However, only one cluster is
visible in the INT images. This cluster is at $z_{\rm phot}=0.50\pm 0.03$. No
more systems are associated with this SZ source.

\begin{figure}
\centering
\includegraphics[width=\columnwidth]{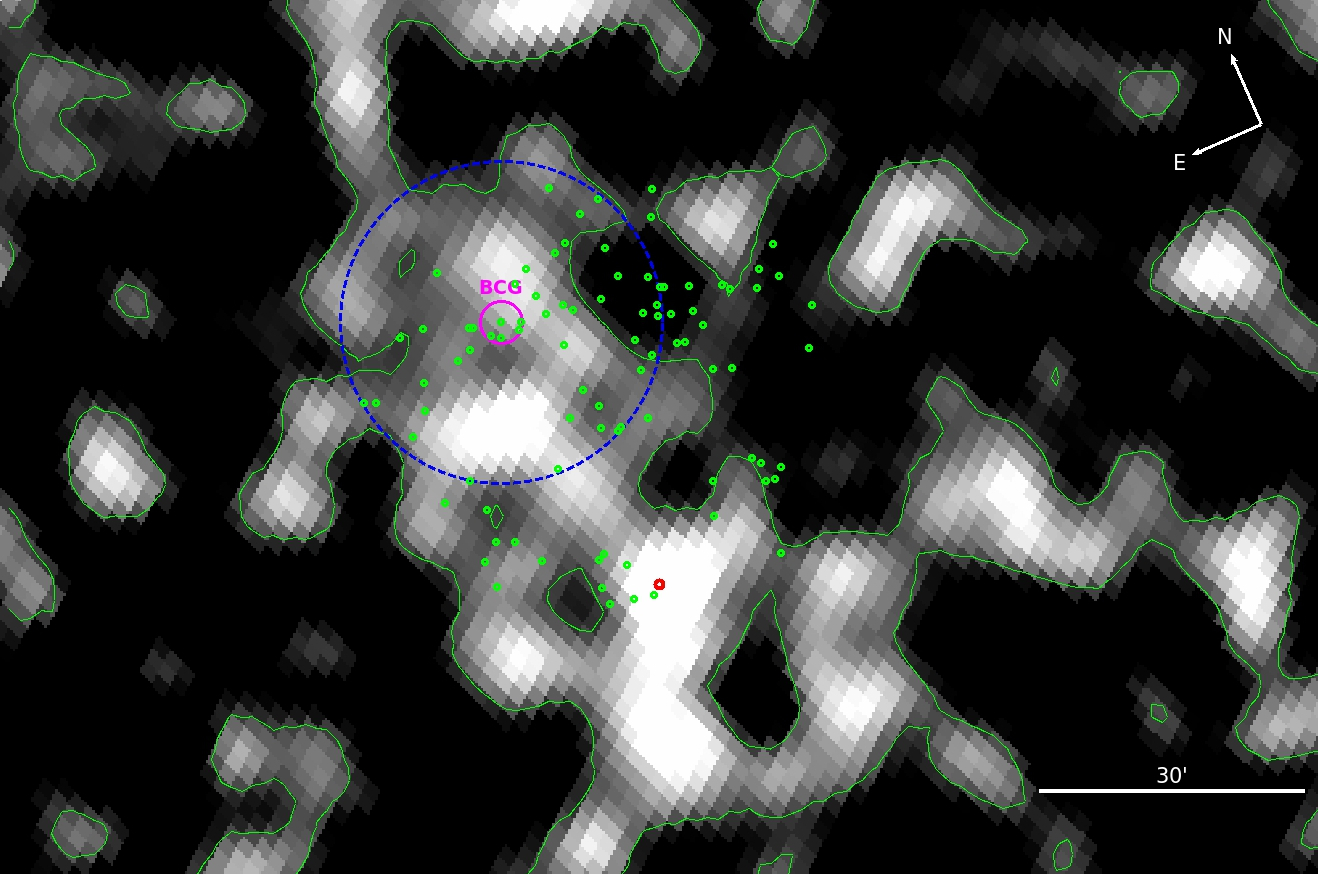}
\caption{SZ emission as seen in the Compton $y$-maps for the source PSZ2
  G328.96$+$71.97. The red dot represents the SZ coordinate as it appears in the
  PSZ2 catalogue. The BCG of the cluster is plotted in magenta, while the rest
  of the galaxies members are shown in green. The blue circle encloses the
  virial radius of this cluster. The 94 cluster members are placed completely
  off the SZ peak and the optical counterpart of this SZ source remains
  unknown.}
\label{Fig1547}
\end{figure}

PSZ2 G328.96$+$71.97 was validated by \cite{Streblyanska18} using SDSS DR12
data. Here, we use new spectroscopic information provided by SDSS DR14 in order
to update the information there reported. We find 94 cluster members at $<z_{\rm
  spec}>=0.090$. However, the BCG of this structure is at $34\parcm6$ from the
\planck pointing. Fig. \ref{Fig1547} shows the scenario around this region. SZ
emission presents a very spread and irregular profile, with several peaks. The
94 cluster members present a $\sigma_{v} \sim 800$\,km\,$s^{-1}$ and a virial
radius of 1.6\,Mpc, but the cluster seems to be placed completely off the SZ
peak. Notice that the distance between the \planck pointing and the optical
cluster centre is larger (double) than the virial radius of the cluster. So, for
all these reasons, we conclude that no optical counterpart is found for this SZ
source, and the actual counterpart (if it exists) is still unknown.
	
\section{Observations of other PSZ2 candidates beyond the LP15 sample}
\label{sec:wise}

In the PSZ2 catalogue, there are 73 clusters validated using the AllWISE
mid-infrared source catalogue \citep{Cutri13}. This catalogue is based on the
observations from the Wide-field Infrared Survey Explorer mission
\citep[WISE,][]{Wright10}. Using the (W1-W2) colour they search for galaxy
over-densities in the redshift range $0.3<z<1.5$. The details of this validation
process can be found in Sect.~7.4 in \cite{PSZ2}. Those objects had {\tt
  validation} $= 16$ in the original PSZ2 catalogue, and thus, they were not
included in the definition of the {\tt LP15} sample.

Here, we present an update on 38 of those ALLWISE sources, providing their
spectroscopic redshifts which were obtained using dedicated observations carried
out with the telescope time within the {\tt LP15}
programme. Table~\ref{tab:wise} presents this information, and it is organised
in the same way as Table~\ref{tab:LP15}.

The double detections PSZ2 G086.28$+$74.76, PSZ2 G139.00$+$50.92 and PSZ2
G141.98$+$69.31 can be considered single detection even if secondary clusters
are detected, because they are very low mass systems, and so not capable of
contributing significantly to the SZ signal.

As noted in earlier by the Planck Collaboration, the number of double cluster detections is relatively high, as compared to other surveys, either in X-rays \citep{PcolintI} or in optical \citep{PcolintXXVI}.

We have cross-checked our sample with two galaxy cluster catalogues WHL
\citep{WHL} and WHY \citep{WHY}, based on optical and infrared data,
respectively. WHL catalogue was published using SDSS data, while WHY used 2MASS
\citep{2MASS}, WISE \cite{Wright10} and SuperCOSMOS \citep{SuperCOSMOS} data. We
only find three matches with WHL. PSZ2 G076.55$+$60.29 and PSZ2 G141.98$+$69.31
will be discussed in the next sub-section, and PSZ2 G021.02$-$29.0 which is also
part of the WHY catalogue. We find seven matches with the WHY catalogue, we are
in a $1\sigma$ agreement in redshift except for PSZ2 G056.38$+$23.36. In this
case, we estimate a photometric redshift of $z_{\rm phot} = 0.21 \pm 0.02$ while
\cite{WHY} reports $z_{\rm phot} = 0.31 \pm 0.04$, compatible within $2\sigma$.
	
We also present in this section an update on four sources that were already
confirmed in the PSZ2 original catalogue \cite{PSZ2} as they were matched with
PSZ1 clusters but without an estimation of their redshifts. Here, we provide the
photometric redshift for three of them and invalidate the already confirmed PSZ2
G198.73$+$13.34, for which we are unable to find any galaxy over-density. In a
future publication, we will discuss this type of sources that we believe are
false validations. These four sources can be found in table~\ref{tab_psz1}.
	
\begin{table*}[t]

\centering
\caption{Update of already known optical counterparts from the PSZ1.}
\tiny
\begin{tabular}{c c c c c c c c c c c}

\noalign{\vskip 3pt\hrule\vskip 5pt}

ID &  Planck Name & SZ SNR & R. A. & Decl.& Dist.($\arcm$) & $z_{\rm phot}$ & $R_{\rm cor}$ & $\sigma_{\rm R}$ & {\tt Flag} & PSZ1 Name \cr

\noalign{\vskip 3pt\hrule\vskip 5pt}

 897 & PSZ2 G196.65$-$45.51 & 4.91 & 03 42 54.40 & $-$08 41 07.70 & 1.52 & 0.25$\pm$0.03 &       $-$    &  $-$ &  2 &  PSZ1 G196.62$-$45.50 \cr 
 901 & PSZ2 G198.73$+$13.34 & 6.03 &     $-$     &      $-$       & $-$  &     $-$       &       $-$    &  $-$ & ND &  PSZ1 G198.67$+$13.34 \cr
1130 & PSZ2 G249.14$+$28.98 & 5.96 & 09 44 57.60 & $-$13 48 11.22 & 1.16 & 0.15$\pm$0.03 & 18.6$\pm$4.3 &  5.3 &  2 &  PSZ1 G249.14$+$28.98 \cr
1539 & PSZ2 G326.73$+$54.80 & 5.92 & 13 45 14.70 & $-$05 32 04.00 & 3.91 & 0.46$\pm$0.05 & 20.3$\pm$4.5 & 10.6 &  2 &  PSZ1 G326.64$+$54.79 \cr

\hline
\end{tabular}

\label{tab_psz1}
\end{table*}

\subsection{Discussion on special cases}
\label{sec:notes_update}	
	
\begin{figure}
\centering
\includegraphics[width=\columnwidth]{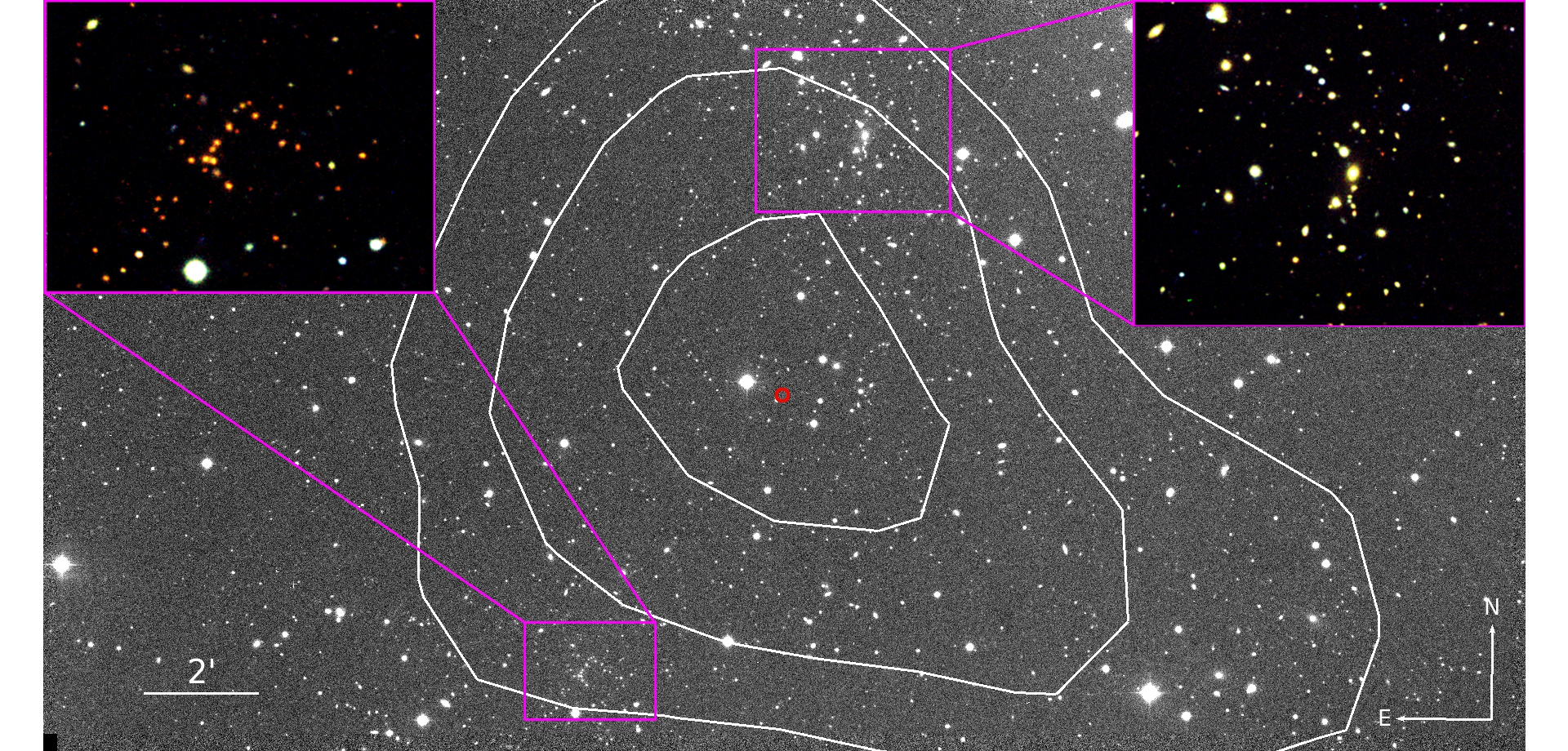}
\caption{Compton $y$-map superimposed to the INT $r'$-band of the PSZ2
  G076.55$+$60.29. White contours correspond to the $3$, $4$ and
  $5\times10^{-6}$ levels of the Compton $y$-map in this area. The \planck
  nominal pointing is represented in red. In the upper corners we show zoomed
  RGB images of the 327-A (left) and 327-B (right), respectively. Both systems,
  at $z_{\rm spec} = 0.632$ and $z_{\rm spec} = 0.287$, are associated with this
  SZ signal. This is a clear example of a multiple detection.}
\label{Fig326}
\end{figure}

We found that PSZ2 G076.55$+$60.29, which it was classified as an individual
counterpart by \cite{Streblyanska18}, is, in fact, a superposition of two
clusters, at $z_{\rm spec}=0.287$ and $z_{\rm spec}=0.632$, respectively. The
first one (327-A) was already proposed as potentially associated cluster. Here,
we confirmed it with 5 spectroscopic members. The distance to the \planck
nominal pointing of the second counterpart (327-B) is slightly greater than
$5\arcm$ but the MILCA $y$-map contours superimposed on an INT image
(Fig. \ref{Fig326}) show that the SZ emission is clearly a superposition of both
clusters. Both counterparts are two of the richest systems studied in this work,
presenting a $\sigma_{R}$ of 23.1 and 8.4, respectively.

PSZ2 G086.28$+$74.76 We find two clusters around the SZ emission at $z_{\rm
  spec} = 0.246$ and $z_{\rm spec} = 0.701$ which we name 381-A and 381-B,
respectively. Both present high-velocity dispersion. However, the centre of
381-A is $8\parcm 93$ away from the \planck centre. So we conclude that this
source has only one optical counterpart at $z_{\rm spec} = 0.701$.

\begin{figure}
\centering
\includegraphics[width=\columnwidth]{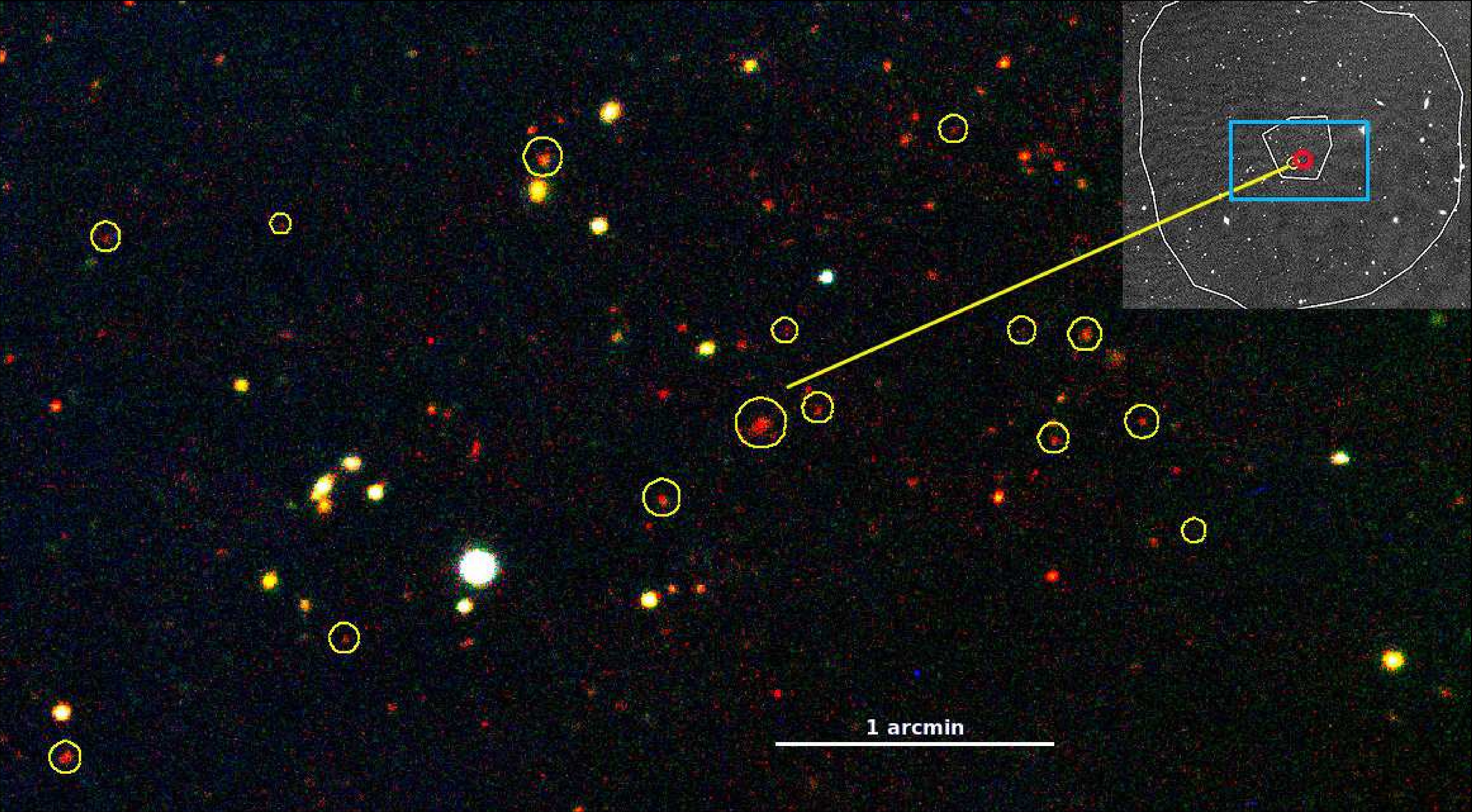}
\caption{The zoomed RGB image of the central area of high-z cluster ($z_{\rm
    spec} = 0.816$) associated with PSZ2 G126.57+51.61. We marked as a yellow
  circles cluster members confirmed spectroscopically. The small top panel shows
  the WFC/INT $i'-$band image with white contours corresponding to the 3 and
  6$\times$10$^{-6}$ levels of the Compton $y$-map in this area. The red circle
  indicates the nominal PSZ2 position. The blue square shows the size of the
  area presented in the main RGB image.}
\label{Fig624}
\end{figure}

PSZ2 G126.57$+$51.61 is one of the most distant cluster in our sample, at
$z_{\rm spec}=0.816$. \citet{Burenin18} published confirmation of one galaxy at
$z_{\rm spec} = 0.815$. This cluster is at the detection limits of our deep
optical images, with most of the members detected almost at the noise level of
the $i'-$band image. No RS for this cluster could be constructed. However,
inspection of the RGB image revealed an over-density of red sources close to the
\planck position supported by the contours extracted from the MILCA $y$-map
(Fig.~\ref{Fig624}).  Our spectroscopic data together with SDSS data confirm
this cluster with 20 members and a $\sigma_{v} \sim 850$\,km\,s$^{-1}$.
	
\begin{figure}
\centering
\includegraphics[width=\columnwidth]{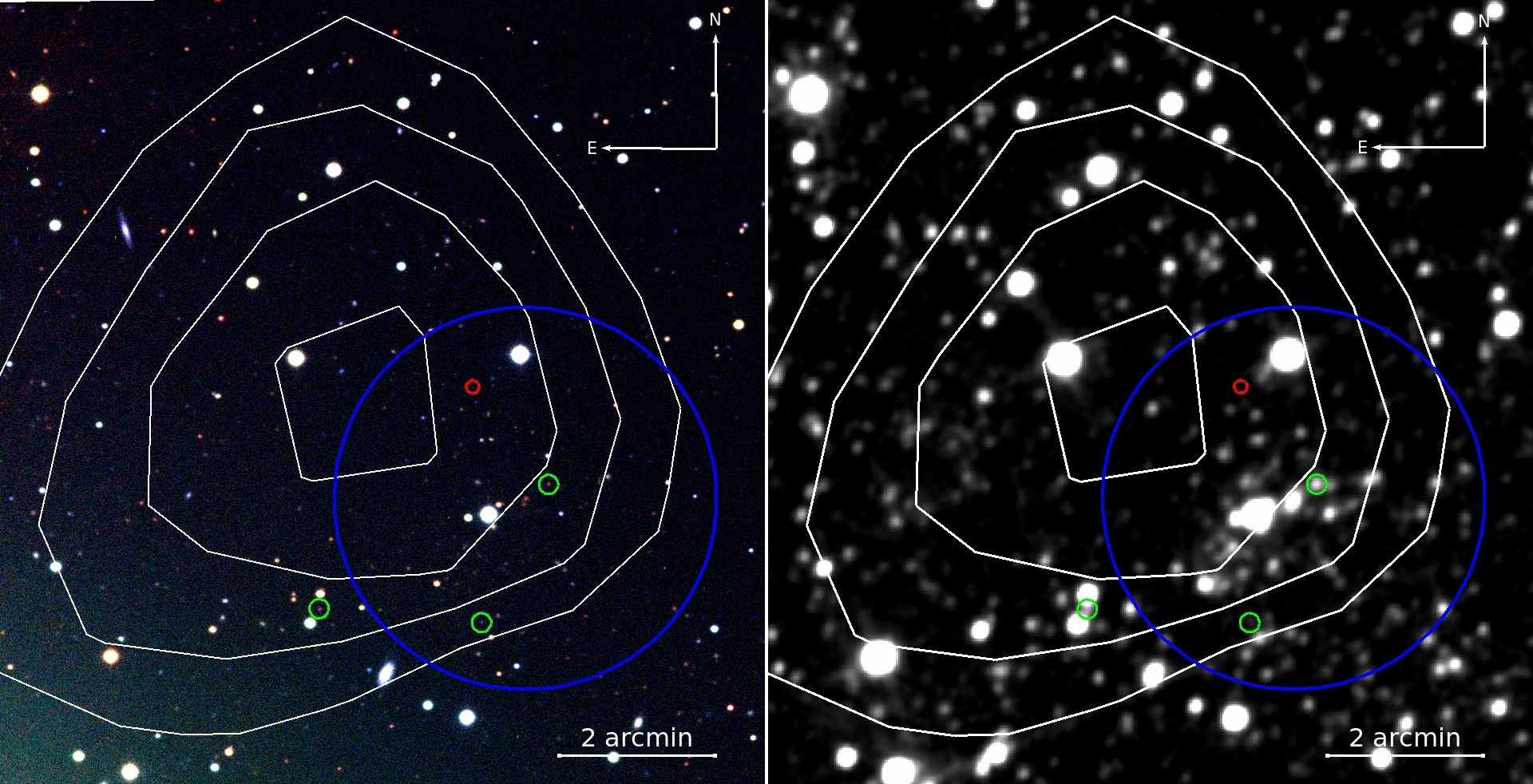}
\caption{Left: RGB image of the source PSZ2 G133.92$-$42.73. Right: WISE
  $W1$-band of the same region. In both images Compton $y$-maps are represented
  in white ($5$, $5.8$, $6.6$ and $7.4\times 10^{-6}$ levels). The blue region
  corresponds to $1$\,Mpc ($2\parcm45$) at the mean redshift of the three
  galaxies represented in green. The \planck nominal pointing is marked in red.}
\label{Fig653}
\end{figure}

PSZ2 G133.92$-$42.73 There is a potential counterpart but it is discarded due to
its low $\sigma_{R}$. Looking at the RGB image (left panel, Fig.~\ref{Fig653})
it seems to be a high redshift cluster as seen in the WISE image (right panel,
Fig. \ref{Fig653}). In SDSS there are three galaxies with $z_{\rm spec} \sim
0.581$ but they are not associated with any galaxy over-density. Deeper imaging
or spectroscopic observations would be needed in order to reject the possibility
of a high-z ($z > 0.8$) cluster.

PSZ2 G139.00$+$50.92 was already confirmed by \cite{Streblyanska18} at $z_{\rm
  phot}=0.6$. We have performed spectroscopic observations for this cluster,
named 681-A in this work, finding a velocity dispersion below the confirmation
limit ($\sigma_{v} < 650$\,km\,s$^{-1}$). We conclude that it is not the main
counterpart to the SZ emission. However, we find another cluster (681-B) at
$z_{\rm spec} = 0.784$ showing a $\sigma_{v} > 800$\,km\,s$^{-1}$, so we
conclude that this last counterpart is the responsible for the SZ emission.

PSZ2 G141.98$+$69.31 This is a case of double detection. We find two
over-densities in this field but once we made spectroscopic observations and
calculate the velocity dispersion, we realised that the object named 690-B
presented a $\sigma_{v} < 400$\,km\,s$^{-1}$, very low to be associated with the
SZ emission. Consequently, we only validate the object proposed in
\cite{Streblyanska18} with a spectroscopic redshift of $z_{\rm spec} = 0.713$,
here named as 690-A.

PSZ2 G270.78$+$36.83 This candidate was already validated by \cite{Streblyanska18}
as a double detection. Here, we spectroscopically confirm one of these
counterparts by detecting 25 cluster members at $z_{\rm spec} = 0.516$, showing
a $\sigma_{v} \sim 900$\,km\,s$^{-1}$. The second counterpart remains
unconfirmed spectroscopically.


\begin{landscape} 

\setlength\LTleft{10pt}
\begin{table}
\caption[]{Updated information of other PSZ2 candidates beyond the LP15 sample.} \label{tab:wise}
\tiny
\begin{tabular}{@{\extracolsep{\fill}}p{1.1cm} c c c c c c c c c c c p{4.5cm} @{}}
\hline \hline 
\noalign{\smallskip}
\,   & \multicolumn{6}{c}{Position (J2000)} \\
\cline{4-5}
\noalign{\smallskip}
ID$^{1}$&  Planck Name & SZ SNR & R. A. & Decl.& Dist.($\arcm$) & $<z_{\rm spec}>$ ; $z_{\rm spec,BCG}$&$N_{\rm spec}$&$z_{\rm phot}$&$R_{\rm cor}$&$\sigma_{\rm R}$&{\tt Flag} & \multicolumn{1}{c}{Notes$^2$} \\

\noalign{\smallskip}
\hline 
\noalign{\smallskip}

  65              & PSZ2 G020.21$-$36.84 & 5.17 & 20 51 57.42 & $-$25 29 17.03 & 1.07 &      $-$       & $-$ & 0.21$\pm$0.02 & 18.1$\pm$4.3 &  3.6 &  2 &  \\
  68$^{a,c}$      & PSZ2 G021.02$-$29.04 & 4.89 & 20 20 28.21 & $-$22 25 14.78 & 1.83 & 0.300 ; 0.3017 & 25  & 0.32$\pm$0.04 &      $-$     &  $-$ &  1 &  WHY/WHL J202028.2$-$222515\\
  92              & PSZ2 G027.77$-$49.72 & 4.60 & 21 53 03.42 & $-$23 34 13.08 & 1.45 & 0.165 ; 0.1641 & 15  & 0.12$\pm$0.02 & 20.6$\pm$4.5 &  4.4 &  1 & 2dF Survey, (1)\\
  93              & PSZ2 G027.81$-$45.93 & 5.24 & 21 37 16.11 & $-$22 32 19.80 & 2.23 &                & $-$ & 0.45$\pm$0.05 & 15.8$\pm$4.0 &  5.2 &  2 &   \\
 120              & PSZ2 G033.83$-$46.57 & 5.53 & 21 45 12.35 & $-$18 42 57.41 & 2.50 &      $-$       & $-$ & 0.32$\pm$0.04 &  3.7$\pm$1.9 &  1.7 &  2 &  WHY J214532.4$-$184130, (2)\\
 206              & PSZ2 G050.98$-$61.48 & 5.10 & 22 58 53.05 & $-$15 35 30.87 & 1.76 &      $-$       & $-$ & 0.18$\pm$0.03 & 14.2$\pm$3.8 &  6.5 &  2 &   \\
 227              & PSZ2 G056.38$+$23.36 & 4.85 & 18 01 16.53 & $+$30 23 20.71 & 3.43 &      $-$       & $-$ & 0.21$\pm$0.02 & 11.4$\pm$3.4 &  7.0 &  2 &  WHY J180116.5$+$302321\\
 277$^{a,c}$      & PSZ2 G066.34$+$26.14 & 5.63 & 18 01 06.52 & $+$39 52 06.73 & 1.64 & 0.622  ; 0.6167 & 52  & 0.63$\pm$0.06 &      $-$     &  $-$ &  1 &   \\
 282              & PSZ2 G066.85$+$22.48 & 4.88 & 18 20 08.12 & $+$39 15 52.93 & 2.05 &      $-$       & $-$ & 0.19$\pm$0.02 & 22.3$\pm$4.7 & 10.3 &  2 &  WHY J182008.2$+$391553\\
 294$^{a,c}$      & PSZ2 G069.39$+$68.05 & 4.51 & 14 21 38.32 & $+$38 21 17.54 & 0.46 & 0.766 ; 0.7630 & 23  & 0.70$\pm$0.06 &      $-$     &  $-$ &  1 &  CIG-J142138.3$+$382118, (3)\\
 304$^{b}$        & PSZ2 G071.82$-$56.55 & 4.78 & 23 09 37.35 & $-$04 09 52.17 & 1.20 & 0.822 ;\hspace{0.4cm}$-$\hspace{0.25cm} & 38  &      $-$      &      $-$     &  $-$ &  1 &   \\
 323              & PSZ2 G075.85$+$15.53 & 4.64 & 19 10 50.62 & $+$44 54 48.99 & 3.30 &      $-$       & $-$ & 0.30$\pm$0.03 &  7.8$\pm$2.8 &  7.3 &  2 &   \\
 327-A$^{a,d}$    & PSZ2 G076.55$+$60.29 & 5.42 & 14 52 00.51 & $+$44 31 21.31 & 4.21 & 0.287 ;\hspace{0.4cm}$-$\hspace{0.25cm} &  5  & 0.25$\pm$0.03 & 35.9$\pm$6.0 & 23.1 &  2 &  WHL J145206.4$+$443235\\
 327-B$^{a}$      &                      &      & 14 52 24.24 & $+$44 22 56.58 & 5.33 & 0.632 ;\hspace{0.4cm}$-$\hspace{0.25cm} &  1  & 0.70$\pm$0.06 & 14.5$\pm$3.8 &  8.4 &  2 & \\
 333              & PSZ2 G078.10$-$83.83 & 4.87 & 00 32 30.94 & $-$22 42 10.95 & 2.00 &      $-$       & $-$ & 0.28$\pm$0.03 & 18.4$\pm$4.3 &  8.7 &  2 &   \\
 355              & PSZ2 G082.37$+$22.35 & 5.93 & 18 44 31.37 & $+$53 00 09.01 & 0.60 &      $-$       & $-$ & 0.29$\pm$0.03 & 14.5$\pm$3.8 &  9.3 &  2 &   \\
 359              & PSZ2 G083.56$+$24.90 & 6.13 & 18 29 28.55 & $+$54 43 08.80 & 1.67 &      $-$       & $-$ & 0.32$\pm$0.04 &  7.0$\pm$2.6 &  2.7 &  2 &   \\
 378              & PSZ2 G085.95$+$25.23 & 5.55 & 18 30 23.81 & $+$56 53 11.12 & 0.62 &      $-$       & $-$ & 0.65$\pm$0.05 &  4.6$\pm$2.1 &  1.5 &  2 &   \\
 381-A            & PSZ2 G086.28$+$74.76 & 5.07 & 13 38 40.43 & $+$38 52 32.57 & 8.93 & 0.246 ;\hspace{0.4cm}$-$\hspace{0.25cm} & 20  &      $-$      &      $-$     &  $-$ &  3 & \\
 381-B$^{a,c}$    &                      &      & 13 37 54.11 & $+$38 53 30.94 & 1.28 & 0.701 ;\hspace{0.4cm}$-$\hspace{0.25cm} & 21  & 0.80$\pm$0.06 &      $-$     &  $-$ &  1 & \\
 394$^{a,c}$      & PSZ2 G087.39$-$34.58 & 4.62 & 22 49 09.53 & $+$19 44 30.50 & 1.93 & 0.772  ;\hspace{0.4cm}$-$\hspace{0.25cm} & 31  & 0.70$\pm$0.07 &      $-$     &  $-$ &  1 &  (4) \\
 468              & PSZ2 G098.75$-$28.63 & 4.74 &     $-$     &        $-$     &  $-$ &      $-$       & $-$ &      $-$      &      $-$     &  $-$ & ND & \\
 475              & PSZ2 G099.55$+$34.23 & 5.34 & 17 10 33.34 & $+$68 44 43.60 & 1.01 &      $-$       & $-$ & 0.31$\pm$0.03 & 15.4$\pm$3.9 &  6.0 &  2 &  WHY J171033.4$+$684443\\
 548              & PSZ2 G113.27$+$48.39 & 5.30 & 13 58 59.49 & $+$67 25 50.29 & 0.75 &      $-$       & $-$ & 0.32$\pm$0.03 &  7.9$\pm$2.8 &  4.5 &  2 &   \\
 581              & PSZ2 G118.49$+$48.17 & 5.16 & 13 23 55.03 & $+$68 39 30.73 & 1.09 &      $-$       & $-$ & 0.35$\pm$0.03 & 24.4$\pm$4.9 &  6.2 &  2 &  WHY J132355.0$+$683931\\
 582              & PSZ2 G118.56$-$13.14 & 4.63 & 00 25 13.35 & $+$49 30 35.84 & 0.50 &      $-$       & $-$ & 0.23$\pm$0.03 & 20.9$\pm$4.6 &  3.7 &  2 &   \\
 590              & PSZ2 G120.30$+$44.47 & 5.31 & 13 16 38.50 & $+$72 32 15.60 & 1.09 &      $-$       & $-$ & 0.23$\pm$0.03 & 23.3$\pm$4.8 &  7.9 &  2 &  WHY J131638.6$+$723217\\
 623$^{b}$        & PSZ2 G126.28$+$65.62 & 4.67 & 12 42 23.33 & $+$51 26 20.98 & 1.67 & 0.819 ; 0.8201 & 16  &      $-$      &       $-$    &  $-$ &  1 & (4)  \\
 625$^{a,c}$      & PSZ2 G126.57$+$51.61 & 6.35 & 12 29 47.56 & $+$65 21 13.41 & 0.33 & 0.817 ;\hspace{0.4cm}$-$\hspace{0.25cm} & 20  & 0.80$\pm$0.10 &       $-$    &  $-$ &  1 & (4) \\
 654$^{a}$        & PSZ2 G133.92$-$42.73 & 4.70 & 01 25 33.36 & $+$19 22 53.51 & 1.65 & 0.581 ;\hspace{0.4cm}$-$\hspace{0.25cm} &  3  & 0.65$\pm$0.07 &  2.3$\pm$1.5 &  1.1 & ND &   \\
 681-A            & PSZ2 G139.00$+$50.92 & 4.98 & 11 20 22.76 & $+$63 14 38.35 & 1.63 & 0.636 ;\hspace{0.4cm}$-$\hspace{0.25cm} & 13  &      $-$      &       $-$    &  $-$ &  3 &   \\
 681-B            &                      &      & 11 20 27.45 & $+$63 14 46.15 & 2.04 & 0.784 ;\hspace{0.4cm}$-$\hspace{0.25cm} & 11  &      $-$      &       $-$    &  $-$ &  1 &   \\
 690-A$^{a,d}$    & PSZ2 G141.98$+$69.31 & 4.71 & 12 12 38.98 & $+$46 21 06.46 & 3.25 & 0.713 ;\hspace{0.4cm}$-$\hspace{0.25cm} & 16  & 0.70$\pm$0.06 &      $-$     &  $-$ &  1 &  WHL J121240.6$+$462123\\
 690-B            &                      &      & 12 12 42.63 & $+$46 21 04.59 & 2.65 & 0.796 ;\hspace{0.4cm}$-$\hspace{0.25cm} &  9  &      $-$      &      $-$     &  $-$ &  3 & \\
 701              & PSZ2 G144.23$-$18.19 & 5.22 & 02 38 55.30 & $+$40 11 11.57 & 0.41 &      $-$       & $-$ & 0.31$\pm$0.03 & 13.4$\pm$3.7 &  2.2 &  2 &   \\
 764              & PSZ2 G159.40$-$40.67 & 5.05 & 02 42 22.99 & $+$14 15 14.60 & 2.81 &      $-$       & $-$ & 0.22$\pm$0.03 & 23.2$\pm$4.8 &  5.4 &  2 &   \\
 768              & PSZ2 G160.83$-$70.63 & 6.30 & 01 39 20.72 & $-$11 22 19.49 & 1.20 &      $-$       & $-$ & 0.24$\pm$0.03 & 26.2$\pm$5.1 &  7.7 &  2 &   \\
 810$^{a,c}$      & PSZ2 G171.08$-$80.38 & 4.90 & 01 21 53.44 & $-$20 33 26.45 & 2.67 & 0.313 ; 0.3134 & 33  & 0.33$\pm$0.03 &      $-$     &  $-$ &  1 &  WHL J012153.4$-$203327\\
 902$^{b}$        & PSZ2 G198.80$-$57.57 & 4.83 & 03 02 06.58 & $-$15 33 41.69 & 0.53 & 0.530 ; 0.5292 & 16  &      $-$      &       $-$    &  $-$ &  1 &   \\
 937$^{b}$        & PSZ2 G208.57$-$44.31 & 4.53 & 04 02 36.08 & $-$15 40 49.56 & 1.47 & 0.820 ; 0.8196 & 17  &      $-$      &       $-$    &  $-$ &  1 &   \\
1254$^{a,c}$      & PSZ2 G270.78$+$36.83 & 4.99 & 11 04 21.06 & $-$19 14 18.34 & 2.55 & 0.516 ; 0.5146 & 25  & 0.52$\pm$0.05 &      $-$     &  $-$ &  1 &  Double detection in (5)\\
1606              & PSZ2 G343.46$+$52.65 & 4.89 & 14 24 23.15 & $-$02 43 49.34 & 0.88 & 0.711 ;\hspace{0.4cm}$-$\hspace{0.25cm} & 20  & 0.70$\pm$0.07 &       $-$    &  $-$ &  1 &  (4) \\

\noalign{\smallskip}
\hline

\end{tabular}
\end{table}

\begin{tablenotes}[flushleft]
\tiny
\item[1] $^1$ SZ targets identified with the ID followed by an A or B label indicate the presence of multiple counterparts.
\item[2] $^2$ References. (1) \cite{2dF}, (2) \cite{Amodeo18}, (3) \cite{Buddendiek15}, (4) \cite{Burenin18}, (5) \cite{Streblyanska18}
\item[a] $^a$ Photometric and/or spectroscopic redshift obtained from SDSS DR14 data.
\item[b] $^b$ No imaging performed, private communication with Saclay group
\item[c] $^c$ Already confirmed in \cite{Streblyanska18}
\item[d] $^d$ Classified as "potentially associated" in \cite{Streblyanska18}

\end{tablenotes}


\end{landscape}

\section{On PSZ2 statistics in the northern sky}
\label{sec:fullpsz2}

In PSZ2 catalogue there are 1003 sources with $Dec.>-15\degree$. After the two
years of {\tt LP15} observations, a total of 226 sources have been observed; 184 of
them were part of the {\tt LP15} sample and thus were not validated at the time the
PSZ2 catalogue was published. In addition, we updated the redshift for 42
additional sources. In this section, we will carry out the statistical analysis
of this ``northern sky'' sub-sample of the PSZ2, such as the purity and effects
that can influence the PSZ detection. For definiteness, we will refer to this
sub-sample as PSZ2-North, which represents the $60\,\%$ of the complete PSZ2
sample.

We note that this PSZ2-North sample also includes some PSZ2 sources associated
with PSZ1 objects, that were observed during the {\tt ITP13}
\citep{paper1,paper2}. There are still five sources ($<0.5\,\%$) that could not
be observed in order to validate the full PSZ2-North, so we exclude them of the
sample for the computation of the statistics in this section.

\begin{figure}
\centering
\includegraphics[width=\columnwidth]{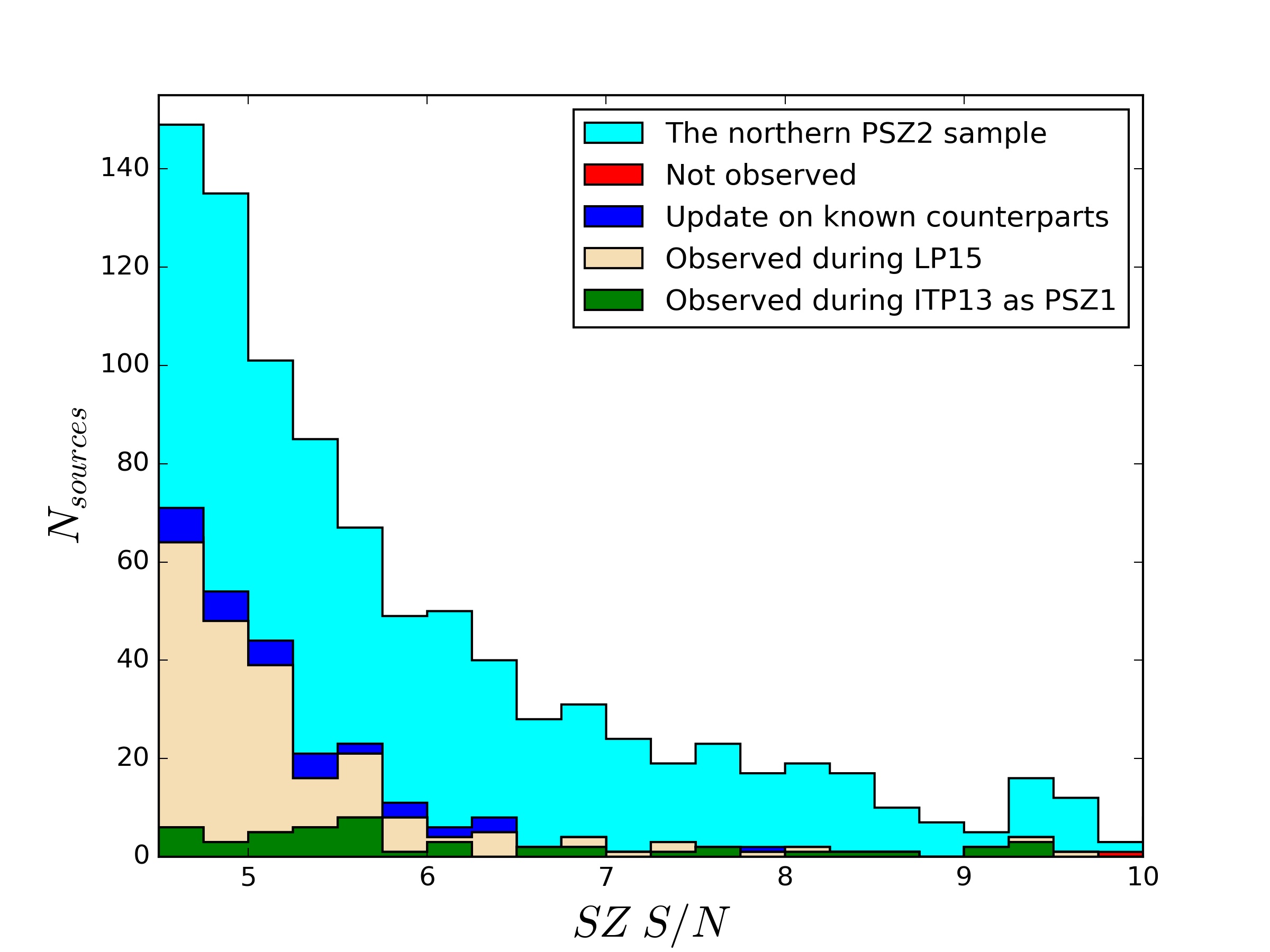}
\caption{PSZ2 cluster counts as a function of the signal-to-noise ratio of the
  SZ detection $S/N$. The PSZ2-North sample is represented in light blue, the
  sources still not observed are represented in red ($<0.5\%$), the updated
  sources described in Sect.~\ref{sec:wise} are shown in dark blue ($3.4\%$) and
  the sources originally not confirmed that were observed during {\tt LP15} and
  {\tt ITP13} are shown in green ($18.4\%$ and $4.8\%$, respectively). The bin
  size is $0.25$.}
\label{FigSNR}
\end{figure}
	
Figure~\ref{FigSNR} shows the number of clusters as a function of the
signal-to-noise ratio in the catalogue. The vast majority of the sources studied
in this work present $S/N < 6$ and it is within this range where this optical
follow-up has performed the largest contribution. In particular, we observed
$37\%$ of the sources with $4.5 < S/N < 6$.

\begin{figure}
\centering
\includegraphics[width=\columnwidth]{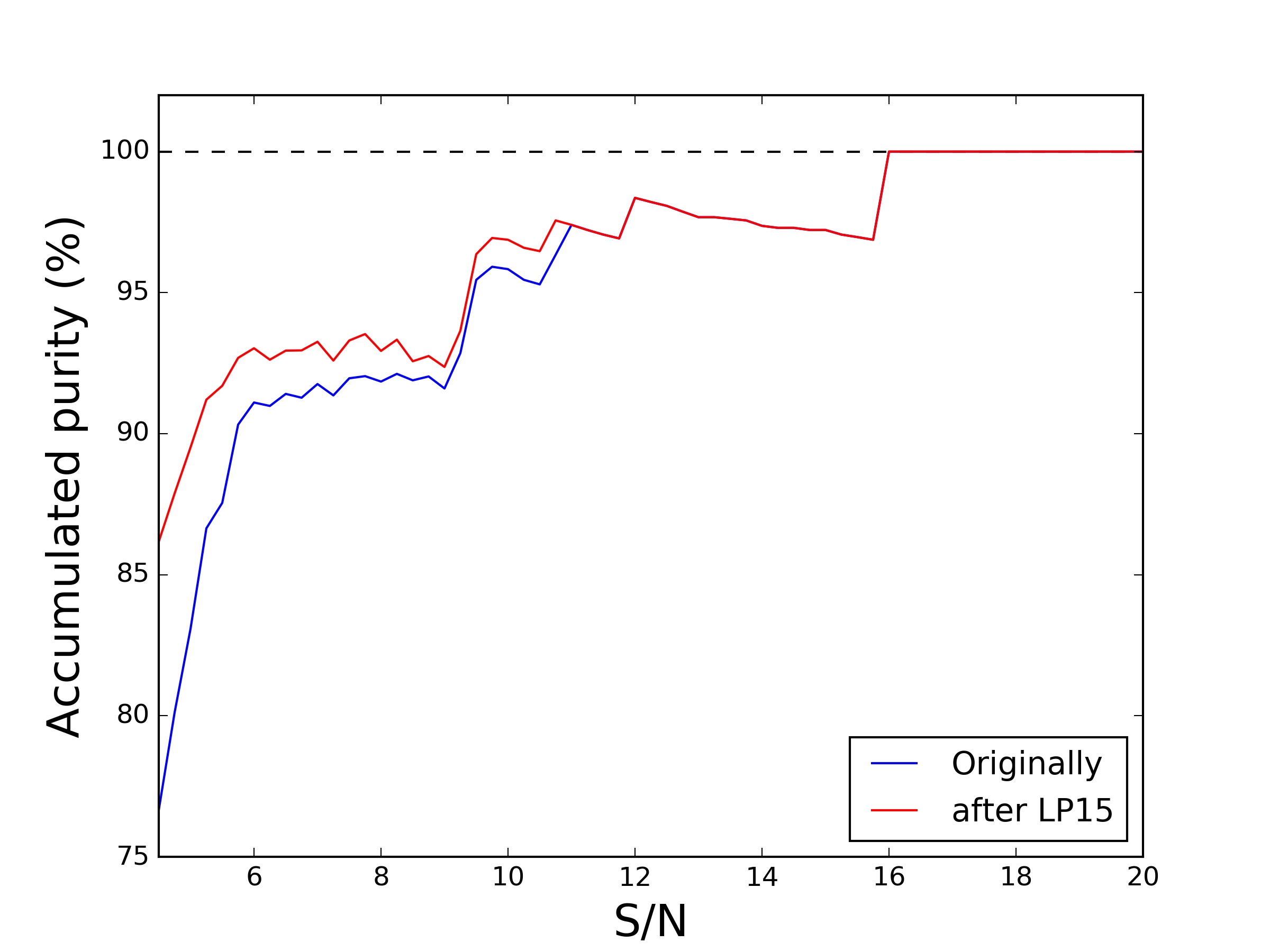}
\caption{Accumulated purity of the PSZ2-North sample (Dec.$>-15\degree$) studied
  as a function of the $S/N$, i.e. the percentage of the sources that are actual
  clusters and related to the SZ signal. In blue we represent the original
  purity of the catalogue and in red the same purity but after this work. }
\label{Figpurity}
\end{figure}

We define the purity as the ratio between confirmed clusters and the total
number of SZ sources. It is important to take into account that we have explored
the optical range in which the dust emission could be masking the possible
counterpart, later in this section we quantify this
effect. Figure~\ref{Figpurity} shows the accumulated purity of the PSZ2-North
sample as a function of the $S/N$. While originally, it showed a purity of
76.7\,\%, after every validation programme to date, the purity increases up to
$86.2\,\%$ for $S/N > 4.5$. The feature in Fig.~\ref{Figpurity} showing a
decrease of purity in the range $12 < S/N < 16$ is due to the existence of one
non-detection listed in the PSZ2 as high $S/N$ source (PSZ2 G153.56+36.82),
studied in detail in Paper I.

\begin{figure}
\centering
\includegraphics[width=\columnwidth]{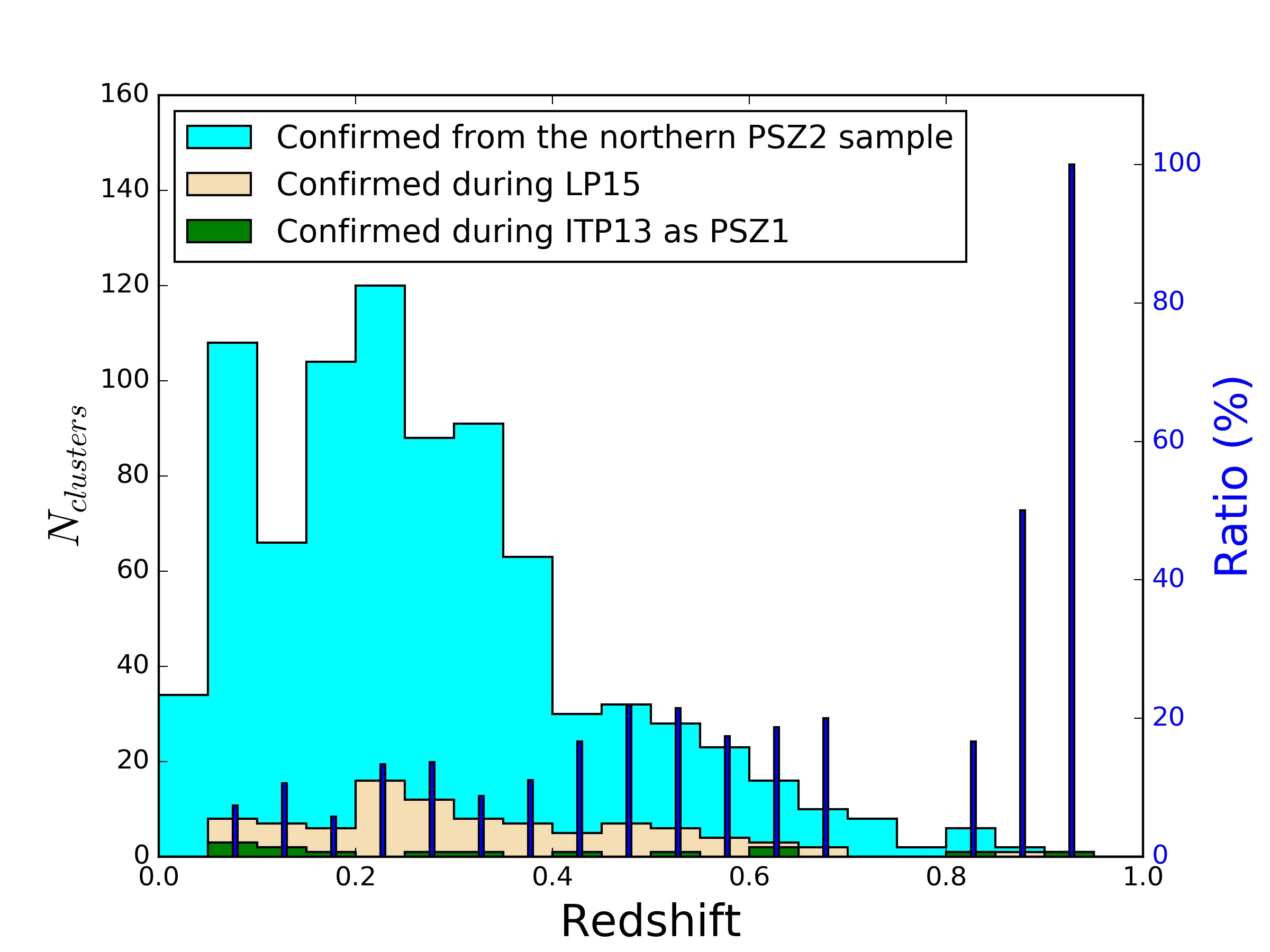}
\caption{Cluster counts as a function of redshift. Colour codes are the same as
  in Fig. \ref{FigSNR}. Dark blue bars represent the ratio between clusters
  confirmed during our follow-ups and the total confirmed clusters. The size of
  the redshift bin is $0.05$.}
\label{FigNz}
\end{figure}
	
Figure~\ref{FigNz} shows the distribution of redshifts of the \planck confirmed
clusters. We note that $77\,\%$ of them present a redshift between $0.05 < z <
0.4$ which is the optimal range for the cluster detection of the \planck
mission. The median redshift of the PSZ2-North sample is 0.23 while the median
redshift of the clusters confirmed during {\tt LP15} is 0.29. While we confirm
about $10\,\%$ of the clusters at $z < 0.4$, this rate is $\sim 20\,\%$ for $z >
0.4$. Moreover, we confirmed in \cite{paper2} the most distant \planck SZ cluster
in the northern hemisphere: the PSZ2 G123.35+25.39, at $z_{\rm phot}=0.95$.

\cite{Burenin17} presented an extension for the PSZ2 catalogue using SDSS and WISE. We find 28 matches between this catalogue and the {\tt LP15} sample. Our results are in good agreement for the majority of the sources. For PSZ2 G069.47$-$29.06 and PSZ2 G130.64$+$37.16, the author only reports one counterpart, while we find two. PSZ2 G069.47-29.06 was discussed in Paper I, where both candidates where confirmed presenting 44 and 30 spectroscopic members, as mentioned in \cite{Zaznobin19}. PSZ2 G130.64$+$37.16 was discussed in Sect. \ref{sec:notes}. On the contrary, for PSZ2 G066.59$-$58.51, we find only one counterpart, while \cite{Burenin17} reports more than one.

We compare our results with those of \cite[][in prep., private
  communication]{Zohren19}. They use the WHT to validate high-z clusters of the
\planck catalogues. They report the redshift, richness and mass for 23
candidates. Twenty of them were also observed during the {\tt LP15}
programme. We agree with their results but for three cases. They claim as well
as \cite{Burenin18} that PSZ2 092.69$+$59.92 has two counterparts, at $z = 0.46$
and $z = 0.84$. Our spectroscopic observations reveal that the galaxy
over-density at $z = 0.84$ is a low mass system as it presents $\sigma_{v} <
450$\,km\,s$^{-1}$. They found PSZ2 G139.00$+$50.92 presents a mass below their
limit for validation. As discussed in Sect. \ref{sec:notes_update}, we find two
possible counterparts, one of them (681-B) showing a $\sigma_{v} >
800$\,km\,s$^{-1}$. PSZ2 G165.41$+$25.93 is also below their mass limit whereas
in our richness analysis it shows a $\sigma_{R} = 1.8$, just above our
validation limit of $\sigma_{R} = 1.5$.

We also compare our results with \cite{Zaznobin19} where the authors report
38 spectroscopic redshift for PSZ2 candidates. We find 20 matches between this catalogue and the {\tt LP15} sample. We find discrepancies in only one case: PSZ2 G202.61$-$26.26. The authors report three spectroscopic redshifts at $z_{\rm spec} = 0.533$ while we find a galaxy over-density at $z_{\rm phot} = 0.23$ but further than $5\arcm$ away from the \planck centre, so not linked to the SZ emission.

\begin{figure}
\centering
\includegraphics[width=\columnwidth]{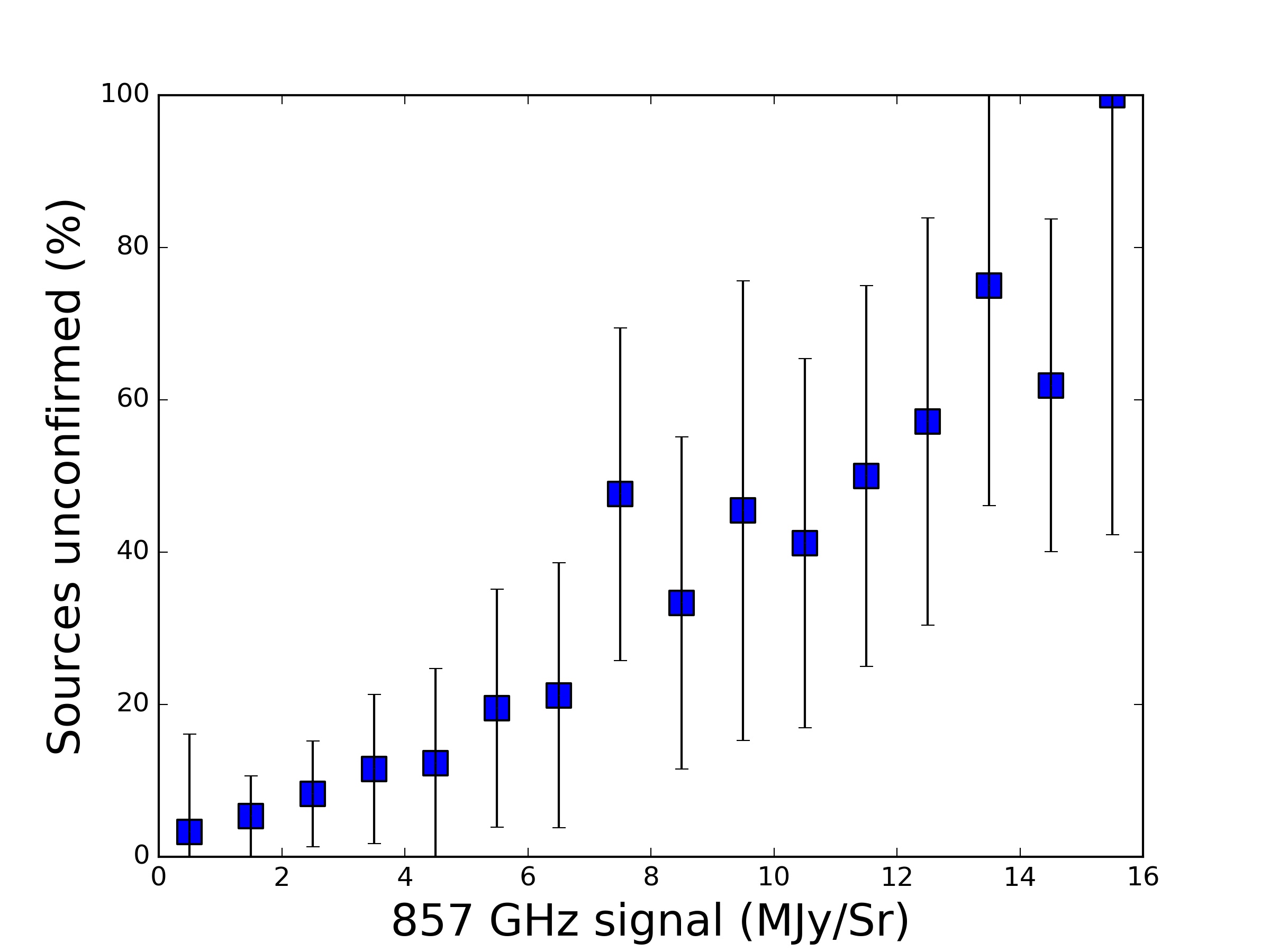}
\caption{Ratio between unconfirmed and total number of sources for the
  PSZ2-North sample (Dec.$>-15\degree$) as a function of the $857$\,GHz signal
  in bins of $1$\,MJy\,sr$^{-1}$. Error bars correspond to a Poisson error in
  the distribution of total number of sources. }
\label{Figdust}
\end{figure}

In order to study the galactic disturbance on the SZ \planck detection, we
compute the number of non-detections as a function of the $857\,GHz$ signal in
the \planck map. This map could be used as a tracer of the thermal dust emission
\citep{2014A&A...571A..11P}. The signal is computed as the mean value within a
region of $0.5\degree$ radius around the nominal pointing in the PSZ2
catalogue. Figure \ref{Figdust} represents the ratio between unconfirmed and
total number of sources for the PSZ2-North sample as a function of the
$857$\,GHz signal in bins of $1$\,MJy\,sr$^{-1}$. This figure shows a clear
correlation between these two magnitudes. Below $7$\,MJy\,sr$^{-1}$ the ratio of
unconfirmed sources is under $20\%$. However, in zones with high dust emission
(mainly places in the galactic plane), the false SZ clusters can be higher than
$60$--$70\%$.
	
\begin{figure}
\centering
\includegraphics[width=\columnwidth]{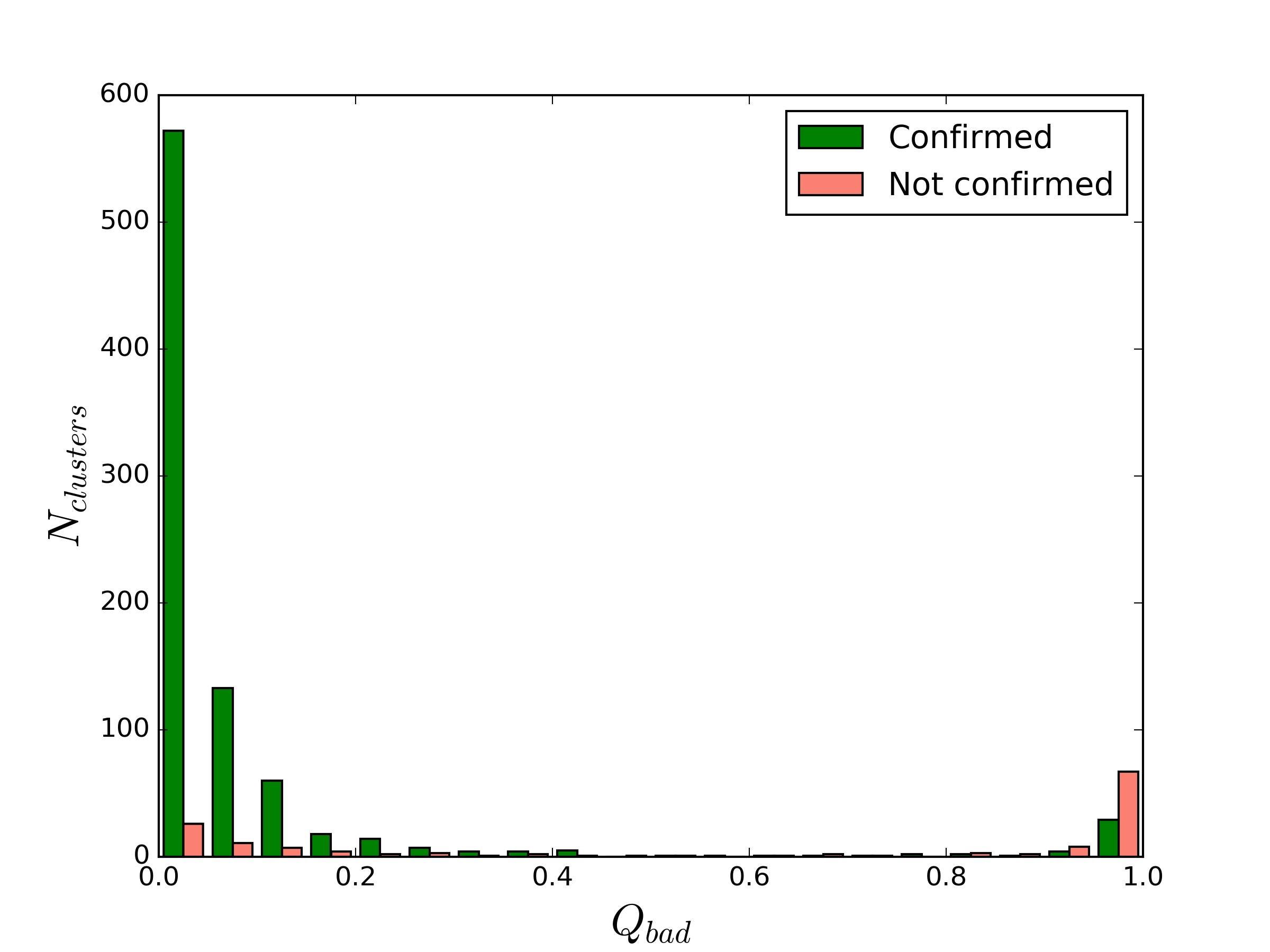}
\caption{Number of cluster-candidates versus the neural network quality flag
  value for the PSZ2-North sample. The confirmed candidates are represented in
  green while the still not confirmed are shown in red. The bin size is 0.05. }
\label{FigQbad}
\end{figure}
	
Figure~\ref{FigQbad} shows the number of cluster-candidates versus the neural
network quality flag value for the PSZ2-North sample ($Q_{bad}$). This value was
defined in \cite{Aghanim15}, it is an indicator of the reliability of a SZ
source to be a real galaxy cluster. Candidates presenting values of $Q_{bad} >
0.6$ are considered low reliable sources. In the PSZ2-North sample, we observe
that the vast majority ($> 93\,\%$) of the clusters with $Q_{bad} < 0.6$ are
actual clusters while less than $33\,\%$ with $Q_{bad} > 0.6$ are confirmed.
		
We also compare the full validation results with \cite{Khatri16} where the
author published a method of validation for the PSZ2 catalogue based on the
combination of CO and $y$-distortion maps. He classifies the sources in five
different groups depending on the value of his estimator: {\it MOC, pMOC, CLG,
  pCLG} and {\it IND}. The signal of the sources classified as {\it MOC} and
{\it pMOC} is considered to come from molecular clouds while {\it CLG} and {\it
  pCLG} come from galaxy clusters. {\it IND} is indeterminable. We find that
$95.2\,\%$ of the {\it IND} and $94.6\,\%$ of the {\it CLG} $+$ {\it pCLG}
correspond to actual validated clusters. On the other hand, $64.7\,\%$ of the
{\it MOC} $+$ {\it pMOC} are also validated clusters. We expected that the
sources with this classification present a lower rate of validation, however, it
is not the case. A possible explanation for these results is that the threshold
used by \cite{Khatri16} to distinguish between molecular clouds and clusters was
shifted towards high values of $\Delta(\Sigma\chi^{2})_{CO-y}$. To illustrate
this fact, for the 59 sources that the author classifies as {\it pMOC}, 48
($81.3\,\%$) are actual clusters. A full study on this matter will be approached
in a future publication where we will discuss the properties of the clusters,
velocity dispersion, masses and their relations with the SZ signal.

\section{Conclusions}
\label{sec:conclusions}

This work presents the final results of the observational programme {\tt LP15}
started in Paper I. We report here about the second year of
observations carried out using INT, TNG and GTC at ORM Observatory, as part of
the optical follow-up confirmation and characterization programme of \planck SZ
sources in the northern hemisphere.
	
During the second year of observations, 78 PSZ2 sources with no known optical
counterpart have been observed. Thanks to a robust confirmation criterion based
on velocity dispersion, when available, and richness estimations we were able to
confirm 40 candidates, providing 18 spectroscopic and 22 photometric redshifts.
	
We update the information on 42 sources that were already validated in the
original PSZ2 catalogue but presented no redshift estimation. We provide
spectroscopic redshift for 20 of them and photometric redshift for 20. We also
perform the richness study and apply the same criteria as for the candidates in
order to check the associations to the SZ signal. We discover that three already
confirmed counterparts were not present in the optical range studied here.
   
At the end of the whole observational programme {\tt LP15}, we were able to
confirm 81 new cluster candidates, with a median redshift of $0.29$ while the
mean redshift of the catalogue is $0.23$. Our main contribution appears in the
redshift interval $0.4 < z < 0.7$, where our confirmations correspond to
$20\,\%$ of the total clusters confirmed in the PSZ2 in that range. The purity of the
catalogue has been updated from $76.7\,\%$ to $86.2\,\%$.
	
Finally, we found a clear correlation between the number of unconfirmed sources
and galactic thermal dust emission. This correlation suggests that there are
spurious detections inside the PSZ2 catalogue. Some of these false detections
have been discussed here. In particular, we find that more than $50\,\%$ of the
sources with a mean signal in the $857$\,GHz maps greater than $7$\,MJy/sr
remain unconfirmed after this work.

\begin{acknowledgements}

This article is based on observations made with a) the Gran Telescopio Canarias
operated by the Instituto de Astrofisica de Canarias, b) the Isaac Newton
Telescope, and the William Herschel Telescope operated by the Isaac Newton Group
of Telescopes, and c) the Italian Telescopio Nazionale Galileo operated by the
Fundacion Galileo Galilei of the INAF (Istituto Nazionale di Astrofisica). All
these facilities are located at the Spanish Roque de los Muchachos Observatory
of the Instituto de Astrofisica de Canarias on the island of La Palma. This
research has been carried out with telescope time awarded for the programme
128-MULTIPLE-16/15B. Also, during our analysis, we used the following databases:
the SZ-Cluster Database operated by the Integrated Data and Operation Center
(IDOC) at the IAS under contract with CNES and CNRS and the Sloan Digital Sky
Survey (SDSS) DR14 database. Funding for the SDSS has been provided by the
Alfred P. Sloan Foundation, the Participating Institutions, the National
Aeronautics and Space Administration, the National Science Foundation, the
U.S. Department of Energy, the Japanese Monbukagakusho, and the Max Planck
Society. This work has been partially funded by the Spanish Ministry of Economy
and Competitiveness (MINECO) under the projects ESP2013-48362-C2-1-P,
AYA2014-60438-P and AYA2017-84185-P. AS and RB acknowledge financial support
from the Spanish Ministry of Economy and Competitiveness (MINECO) under the
Severo Ochoa Programmes SEV-2011-0187 and SEV-2015-0548. HL is funded by PUT1627
grant from the Estonian Research Council and by the European Structural Funds
grant for the Centre of Excellence "Dark Matter in (Astro)particle Physics and
Cosmology" TK133. Some of the results in this paper have been derived using the
{\sc HEALPix} \citep{Healpix} package.

\end{acknowledgements}

\bibliographystyle{aa} 
\bibliography{LP15_year2.bib}

\end{document}